\documentclass[preprint,10pt,authoryear]{elsarticle}
\usepackage{amsmath}
\usepackage{bm}
\usepackage{amsthm}
\usepackage{natbib}
\usepackage{adjustbox}
\usepackage{float}
\usepackage{hyperref}
\usepackage{multirow}

\usepackage{subcaption}
\usepackage{algorithm2e}
\newtheorem{definition}{Definition}

\RestyleAlgo{ruled}
\usepackage{graphicx}
\usepackage{orcidlink}

\usepackage{xcolor}

\usepackage{amssymb}

\journal{arXiv}

\usepackage[a4paper, total={6in, 8in}]{geometry}

\begin{document}

\begin{frontmatter}



\title{Mean regression for (0,1) responses via beta scale mixtures}


\author[inst1]{Otto, Arno \corref{cor1}\orcidlink{0000-0002-6565-2910}}
\ead{arno.otto@up.ac.za}
\cortext[cor1]{Corresponding author}
\author[inst1]{Bekker, Andri\"ette \orcidlink{0000-0003-4793-5674}}
\author[inst3]{Ferreira, Johan T.\orcidlink{0000-0002-5945-6550}}
\author[inst1]{Rathebe, Lebogang}

\affiliation[inst1]{organization={Department of Statistics, University of Pretoria, Pretoria, South Africa}}
\affiliation[inst3]{organization={School of Statistics and Actuarial Science, University of the Witwatersrand, Johannesburg, South Africa}}

\begin{abstract}

To achieve a greater general flexibility for modeling heavy-tailed bounded responses, a beta scale mixture model is proposed.  Each member of the family is obtained by multiplying the scale parameter of the conditional beta distribution by a mixing random variable taking values on all or part of the positive real line and whose distribution depends on a single parameter governing the tail behavior of the resulting compound distribution. These family members allow for a wider range of values for skewness and kurtosis. To validate the effectiveness of the proposed model, we conduct experiments on both simulated data and real datasets. The results indicate that the beta scale mixture model demonstrates superior performance relative to the classical beta regression model and alternative competing methods for modeling responses on the bounded unit domain.
\end{abstract}


\end{frontmatter}

\section{Introduction} \label{sec:Intro}
Modeling continuous responses constrained to the unit interval $(0,1)$ is a recurring challenge in many applied fields, including medicine \citep{guolo2014beta, hunger2012longitudinal}, natural sciences \citep{geissinger2022case}, and environmental studies \citep{douma2019analysing}. Beta regression has become a standard tool in this setting due to its ability to accommodate a variety of distributional shapes on the unit interval. In its classical formulation, beta regression is typically parameterized in terms of a mean and a dispersion (or precision) parameter, allowing the conditional mean of the response to be linked to covariates through a regression structure \citep{ferrari2004beta, smithson2006better}. 

Despite its widespread use, classical beta regression can be sensitive to departures from modeling assumptions. The model remains limited in its ability to accommodate a wide range of phenomena, particularly those exhibiting heavy-tailed characteristics; for example, real-world data are often contaminated with atypical observations. Following  \cite{barnett1994outliers} and \cite{bayes2012new}, we refer to an atypical value as an observation (or set of observations) that appears to be inconsistent with the rest of the data. The lack of {robustness} of classical beta regression to atypical observations has been widely documented in the literature \citep{ghosh2019robust, niekerk2019beta}. Several authors have noted that standard likelihood-based inference under the mean–dispersion parameterization can be heavily affected by atypical observations. Such features can have a disproportionate influence on likelihood-based inference under mean–dispersion parameterizations, leading to biased estimates and reduced predictive performance. Existing contributions range from classical approaches focused on identifying or downweighting influential observations prior to inference, to Bayesian and mixture-based formulations designed to explicitly accommodate contamination or deviations from standard assumptions \citep[e.g.,][]{espinheira2008influence, espinheira2008beta, bayes2012new,  migliorati2018new, ghosh2019robust,niekerk2019beta,di2020robustness}. Therefore, the existing body of work consistently reveals that classical beta regression models lack robustness to outliers and are ill-suited for handling heavy-tailed behavior, leaving this issue as a central and unresolved problem in the modeling of bounded response variables. 

Another issue arises when modeling higher-order moments with the beta distribution. Although the skewness and kurtosis of the beta distribution can, in principle, take on a wide range of values as 
its two parameters vary, this does not mean that we have direct control over them \citep{di2020robustness}. To clarify, suppose we use the method of moments to estimate 
its parameters by matching the sample mean and variance to the corresponding moments of the beta distribution. Once 
they are determined in this way, the skewness and kurtosis are fully implied by that parameter pair and cannot be adjusted independently. Consequently, even though the beta distribution allows for substantial variation in higher-order moments across its parameter space, for any fixed mean–variance combination, the implied skewness and kurtosis are fixed, and there is no guarantee that they will match the empirical higher-order moments observed in the data. See \cite{jones2015families}, \cite{arellano2013centred} and \cite{otto2025contaminated,otto2026modeling} for similar arguments. 


To overcome these limitations, we enhance model flexibility by introducing a beta scale mixture (BSM) model for bounded responses on the unit interval.
The underlying idea is inspired from one of the most famous compound models, the normal scale mixture (see \citet{andrews1974scale} and \citet{watanabe2003algorithm}), in which the variability-related parameter is scaled by a convenient random variable, and has been extended to various different cases (\citealp{boris2008scale, punzo2021modeling, otto2024alternative}).
Following from \citet{andrews1974scale}, given a specified reference model, $f(y;w)$, a scale
mixture model is constructed by superimposing a mixing random variable $W$ with mixing probability distribution $h(w)$, on the variability parameter of the conditional (base) model. In this way, we can conveniently obtain the stochastic representation, which is useful for generating random numbers for the BSM model. The resulting model's flexibility facilitates greater control over the impact of heavy-tailed phenomena.
In this work, we focus on four mixing distributions -- the Bernoulli, gamma, lognormal, and inverse Gaussian -- to demonstrate the flexibility of the BSM distribution. The resulting models exhibit a wider range of skewness and kurtosis than the reference beta distribution, thereby better accommodating heavy-tailedness. Moreover, the BSM framework retains the attractive property of interpretable parameters, facilitating meaningful inference in practical applications \citep{ley2015flexible,azzalini2022overview}.



Our contributions are summarized as follows: In Section \ref{Sec:BSM}, we formulate the BSM model with a tractable hierarchical structure and interpretable parameters, and in Section \ref{sec: Examples of UBSM}, we present several illustrative member distributions. Parameter estimation via maximum likelihood is discussed in Section \ref{sec:Maximum likelihood estimation}. In Section \ref{sec: Sensitivity analysis}, we conduct a sensitivity analysis to assess the impact of atypical observations on the estimators. Finally, in Section \ref{sec:Data application}, we demonstrate the practical relevance of the proposed BSM regression model using real-world data, evaluating its performance on heavy-tailed data using the AIC and BIC criteria. The results show that the proposed models provide a substantial improvement over the classical beta regression model. Section \ref{Sec:conclusion} concludes the paper.

\section{Beta scale mixture}\label{Sec:BSM}
\subsection{Preliminaries: the mean parameterized beta distribution} \label{sec:UB DIST}
The beta distribution is commonly used to model data on the unit interval, such as rates, proportions, and probabilities. To facilitate its application in a regression context, \cite{ferrari2004beta} introduced the mean-parameterized beta distribution, which expresses the distribution in terms of its mean and a variability parameter.   
Specifically, consider a random variable $Y$ is said to have the beta (B) distribution if its probability distribution function (PDF) is given by
\begin{equation}\label{eq: pdf UB}
f_{\text{B}} (y; \mu, \phi) = \frac{y^{\frac{\mu}{\phi}-1}(1-y)^{\frac{1-\mu}{\phi}-1}}{\mathrm{B}\left({\frac{\mu}{\phi}},{\frac{1-\mu}{\phi}}\right)} ,\quad 0< y<1,
\end{equation}
where $\mu$ $\in$ $(0,1)$ denotes the mean, $\phi >0$ is a variability parameter, and $\mathrm{B}\left(\cdot,\cdot\right)$ is the beta function. If $Y$ had the PDF in \eqref{eq: pdf UB}, we denote it as $Y\sim \mathcal{B}(\mu,\phi)$,
 The mean-parameterized beta distribution in \eqref{eq: pdf UB} is directly related to the classical beta distribution through the transformation $\alpha =\frac{\mu}{\phi}$ and $\beta =\frac{1-\mu}{\phi}$.
\begin{figure}[h!]
	\centering
	\begin{subfigure}[h]{0.48\textwidth}
		\centering
  \includegraphics[scale=0.48]{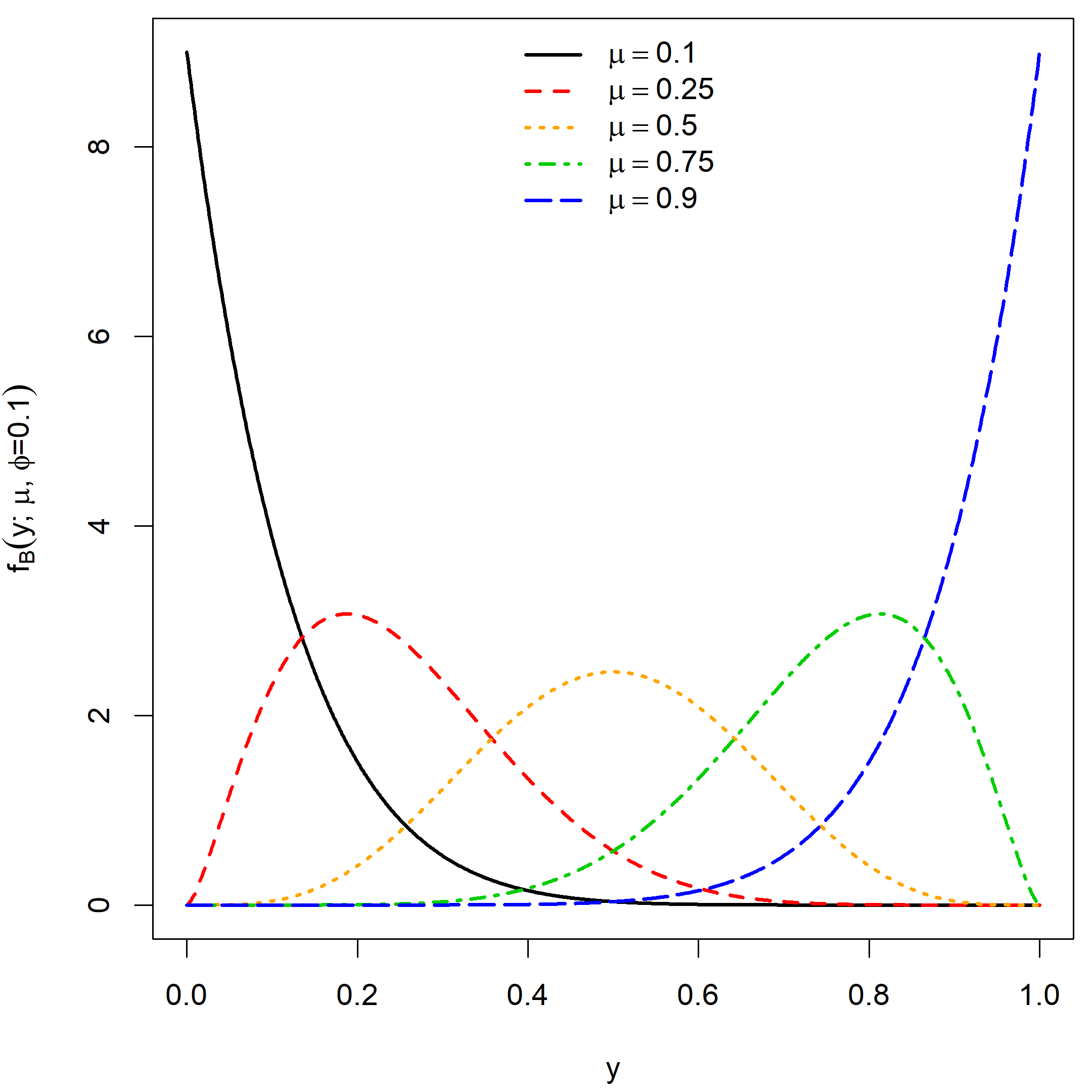}
		\caption{Varying values of $\mu$, when $\phi=0.1$  }
	\end{subfigure}
 \begin{subfigure}[h]{0.48\textwidth}
		\centering
		\includegraphics[scale=0.48]{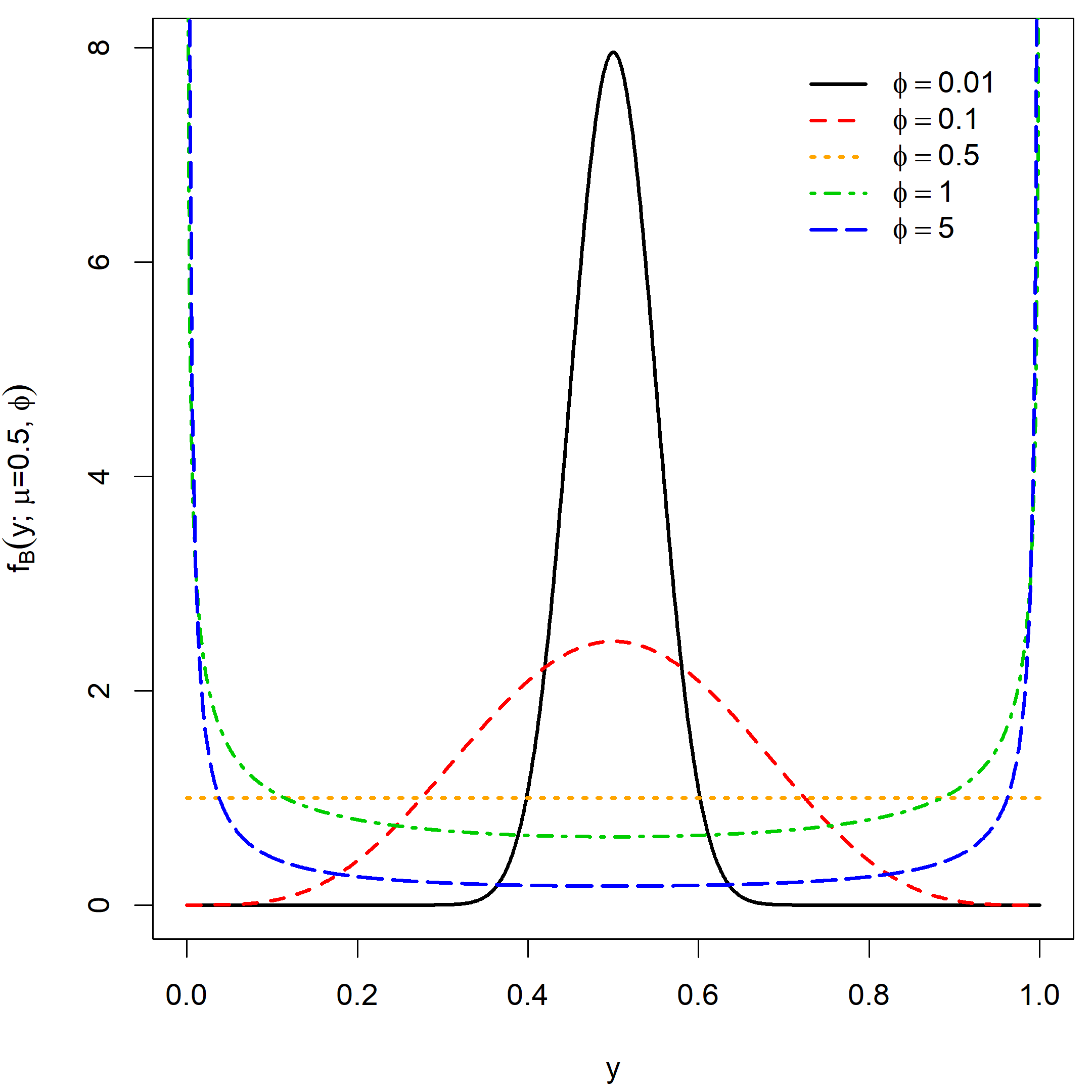}
		\caption{Varying values of $\phi$, when $\mu=0.5$  }
	\end{subfigure}
	\caption{Plots of the beta distribution \eqref{eq: pdf UB} for varying parameter values.}
\label{plot beta}
\end{figure}

 The flexibility of the beta distribution is illustrated in \figurename~\ref{plot beta}, which demonstrates the wide range of distributional shapes it can exhibit, including unimodal, bimodal, uniform, J-shaped, and reverse J-shaped forms. The effect of the mean parameter $\mu$ on skewness is clearly observed: the distribution is right-skewed for $\mu<0.5$, symmetric when $\mu=0.5$, and left-skewed for 
$\mu>0.5$.
 
 The moments, or shape characteristics, of practical interest of $Y\sim B(\mu,\phi)$ are:
\begin{align*}
 \mathrm{E}_{\text{B}}(Y) = \mu,
 \end{align*}

\begin{align*}\label{eq UB var}
\text{Var}_{\text{B}}(Y) = \frac{\mu(1-\mu)\phi}{1+\phi},
\end{align*}

\begin{align*}
    \text{Skewness}_{\text{B}}(Y)=\frac{2(1-2\mu)\sqrt{1+\frac{1}{\phi}}}{\left(2+\frac{1}{\phi}\right)\sqrt{\mu(1-\mu)}},
\end{align*}
\begin{align*}
\text{ExcessKurtosis}_{\text{B}}(Y) = \frac{6\phi(1+\phi-\mu(1-\mu)(5+6\phi))}{\mu(1-\mu)(1+2\phi)(1+3\phi)}.    
\end{align*}

\subsection{Beta scale mixtures}\label{Chapter : Intro UBSM}
This section proposes a framework for reweighting the tails of the beta distribution. The appeal of scale mixture distributions lies in their enhanced flexibility, which allows them to even accommodate potential atypical values. This flexibility is achieved by scaling the variability parameter with a mixing random variable that takes values on all or part of the positive real line. The distribution of this mixing variable depends on a parameter vector $\boldsymbol{\theta}$, which governs the tail behaviour of the resulting model.

\begin{definition}
A random variable $Y$ is said to have the beta scale mixture (BSM) distribution with mean $\mu \in (0,1)$, variability parameter $\phi>0$, and tail weight $\boldsymbol{\theta}$, if the PDF is given by:
\begin{equation}
 \begin{aligned}\label{eq:2}
    f_{\text{BSM}}(y;\mu,\phi,\boldsymbol{\theta}) &= \int_{S_{h}} f_{\text{B}}(y;\mu,\phi/w)h(w;\boldsymbol{\theta}) \,dw , \quad 0<y<1, \\
&= \sum_{w} f_{\text{B}}(y;\mu,\phi/w)h(w;\boldsymbol{\theta}),
\end{aligned}
\end{equation}
   where $h(w;\boldsymbol{\theta})$ is the mixing probability distribution, with $S_{h}\subseteq(0,\infty)$, depending on the vector of parameters $\boldsymbol{\theta}$.  ${Y}$ is then denoted by $Y\sim \mathcal{BSM}(\mu,\phi,\boldsymbol{\theta})$, and can be hierarchically defined as:
\begin{equation}\label{prop:1}
    \begin{aligned}
     Y|W=w &\sim \mathcal{B}\left(\mu,\frac{\phi}{w}\right),\\
     W  &\sim h(\boldsymbol{\theta}).
\end{aligned}
\end{equation}
\end{definition}

A random variable with a BSM distribution can be thought of as a composite or compound distribution with the same location $\mu$, but with a different scale $\phi/w$. The component beta distributions are not taken uniformly from the set, but according to a set of weights determined by the probabilistic behaviour of $W$. Note that if $W$ is degenerate in 1 (i.e., $W\equiv1$ which implies that $P(W=1)=1$) then $\mathcal{B}(\mu,\phi)$ in \eqref{eq: pdf UB} is obtained.

It is easy to show that the $k^{\text{th}}$ moment of $Y\sim \mathcal{BSM}(\mu,\phi,\boldsymbol{\theta})$  is given by
\begin{equation}
    \begin{aligned}\label{prop:2}
        &\mathrm{E}_{\text{BSM}}\left(Y^k\right)= \mathrm{E}_{h}\left[\mathrm{E}_{\text{B}}\left(Y^k|W=w\right)\right],
        \end{aligned}
\end{equation}
where the subscripts of the expected value depict the distribution used to compute the expectations.
The moments of the BSM distribution are given below, using the hierarchical structure in (\ref{prop:1}). The mean, variance, skewness and kurtosis of $Y\sim \mathcal{BSM}(\mu,\phi,\boldsymbol{\theta})$ is:
\begin{equation*}\label{UBSM:Mean}
    \begin{aligned}
        &\mathrm{E}_{\text{BSM}}(Y)= \mu,
        \end{aligned}
\end{equation*}

\begin{align*}\label{UBSM:VAR}
    \text{Var}_{\text{BSM}}(Y)  = \mathrm{E}_{h}\left[\mathrm{Var}_{\text{B}}(Y|W=w)\right]= \mu(1-\mu)\phi\mathrm{E}_h\left(\frac{1}{\phi+W}\right),
\end{align*}
\begin{align*}
    \mathrm{Skew}_{\text{BSM}}(Y)=\frac{2(1-2\mu)\mathrm{E}_h\left(\frac{1}{(\phi+W)(2\phi+W)}\right)}{\sqrt{\mu(1-\mu)\phi \mathrm{E}_h\left(\frac{1}{\phi+W}\right)^3}},
\end{align*}

\scalebox{0.8}{%
$\begin{aligned}
\mathrm{ExcessKurt}_{\text{BSM}}(Y)&=\frac{3\left[(7\mu^2-7\mu+2)\mathrm{E}_h\left(\frac{1}{\phi+W}\right)-4(1-2\mu)^2\mathrm{E}_h\left(\frac{1}{2\phi+W}\right)+(3\mu-1)(3\mu-2)
\mathrm{E}_h\left(\frac{1}{3\phi+W}\right)+\mu(\mu-1)\phi\mathrm{E}_h
\left(\frac{1}{\phi+W}\right)^2\right]}{2\mu(1-\mu)\phi\mathrm{E}_h\left(\frac{1}{\phi+W}\right)}.
\end{aligned}
$
}


\subsection{The beta scale mixture regression model}
A regression model based on the BSM distribution in \eqref{eq:2} follows by conditioning the distribution of the responses $Y_i$, $i=1,\dots,n$, on a $k$-dimensional vector of covariates. 
Let $\boldsymbol{x}$ represent possible values of the covariates $\boldsymbol{X}$, which have dimension $k$. The BSM regression model is then specified through the following link function
\begin{align*}
    g(\mu(\boldsymbol{x};\boldsymbol{\beta}))=\mathrm{logit}(\mu(\boldsymbol{x};\boldsymbol{\beta}))=\tilde{\boldsymbol{x}} '\boldsymbol{\beta},
\end{align*}
where $\boldsymbol{\beta}= (\beta_0,\beta_1,\dots,\beta_k)$ is a vector of unknown regression coeficients, $\tilde{\boldsymbol{x}}=(1,\boldsymbol{x}')'$ accounts for the intercept. The considered link function, even if the most commonly used, is only an example of possible functions that can be considered. The inverse of the link function leads to 
\begin{align*}
    \mu(\boldsymbol{x};\boldsymbol{\beta})=g^{-1}(\tilde{\boldsymbol{x}} '\boldsymbol{\beta})=\frac{\mathrm{e^{\tilde{\boldsymbol{x}} '\boldsymbol{\beta}}}}{1+e^{\tilde{\boldsymbol{x}} '\boldsymbol{\beta}}}.
\end{align*}
The conditional distribution of $Y$ according to the BSM regression model can be written as
\begin{align*}\
Y|\boldsymbol{X}=\boldsymbol{x}\sim \mathcal{BSM}(\mu(\boldsymbol{x};\boldsymbol{\beta}),\phi, \boldsymbol{\theta}).
\end{align*}

\section{Examples of the beta scale mixtures} \label{sec: Examples of UBSM}
In this section, examples of the BSM model are presented by selecting different mixing probability 
density (or mass) functions in each case. 
With the exception of the two-point beta (TPB) distribution, simple closed-form expressions of the PDFs for the BSM family are not analytically tractable. To observe the behaviour and shape of the distributions and to understand the impact of varying the parameters, numerical evaluation is necessary.

Because the beta distribution can be bimodal, the BSM distribution can likewise be unimodal, bimodal, or even trimodal. This
flexibility allows the BSM distribution to model W-shaped data, where data is clustered at both tails, like the U-shape of the
beta distribution, while retaining an additional central mode (\citealp{gallop2013model}, \citealp{keller2022w}, and \citealp{otto2026modeling}).

\subsection{Two-point beta distribution}\label{Bernoulli unimodal beta }
\begin{figure}[h!]
	\centering
	\begin{subfigure}[h]{0.48\textwidth}
		\centering
  \includegraphics[scale=0.48]{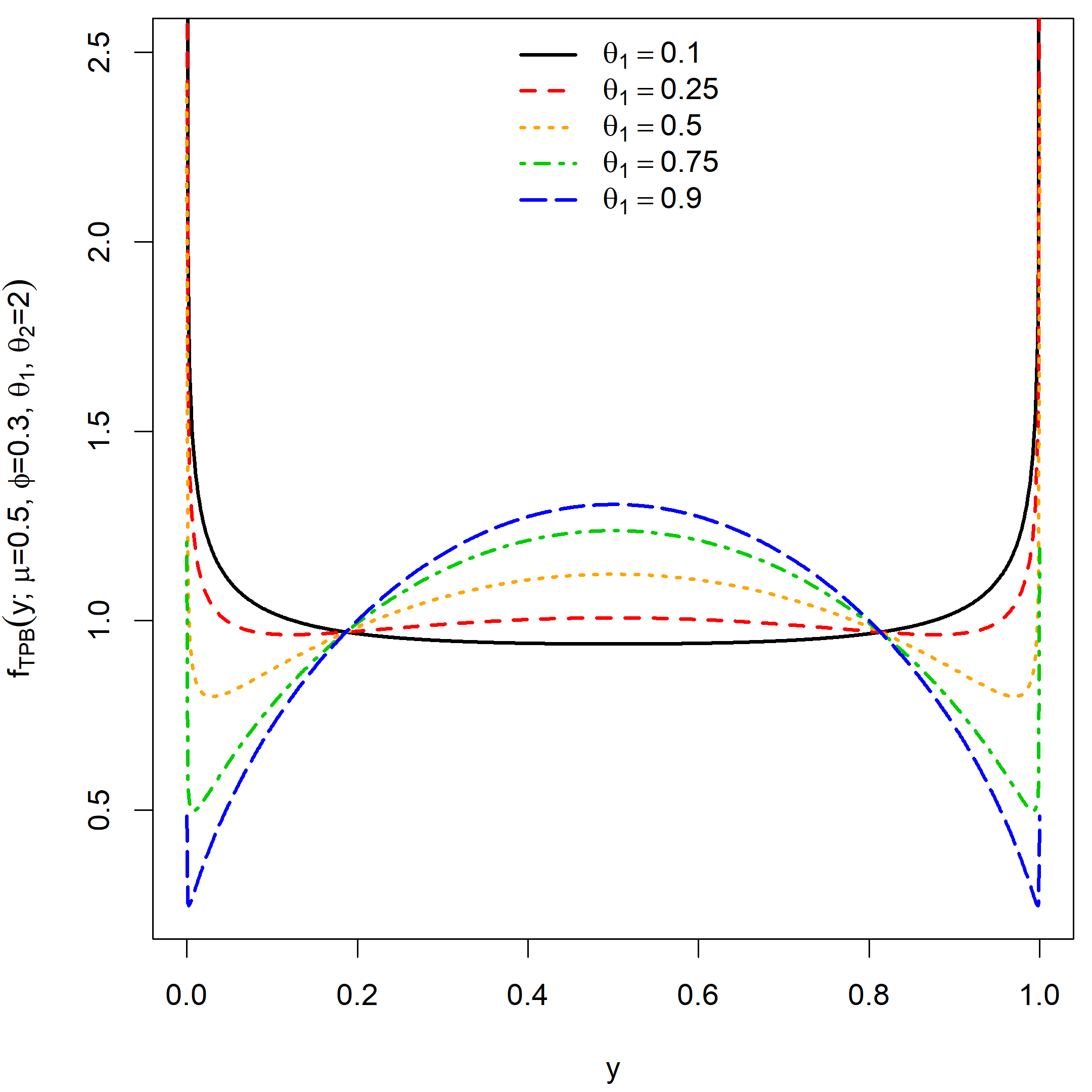}
		\caption{Varying values of $\theta_1$, when $\mu=0.5, \phi=0.3, \theta_2=2$.  }
	\end{subfigure}
 \begin{subfigure}[h]{0.48\textwidth}
		\centering
		\includegraphics[scale=0.48]{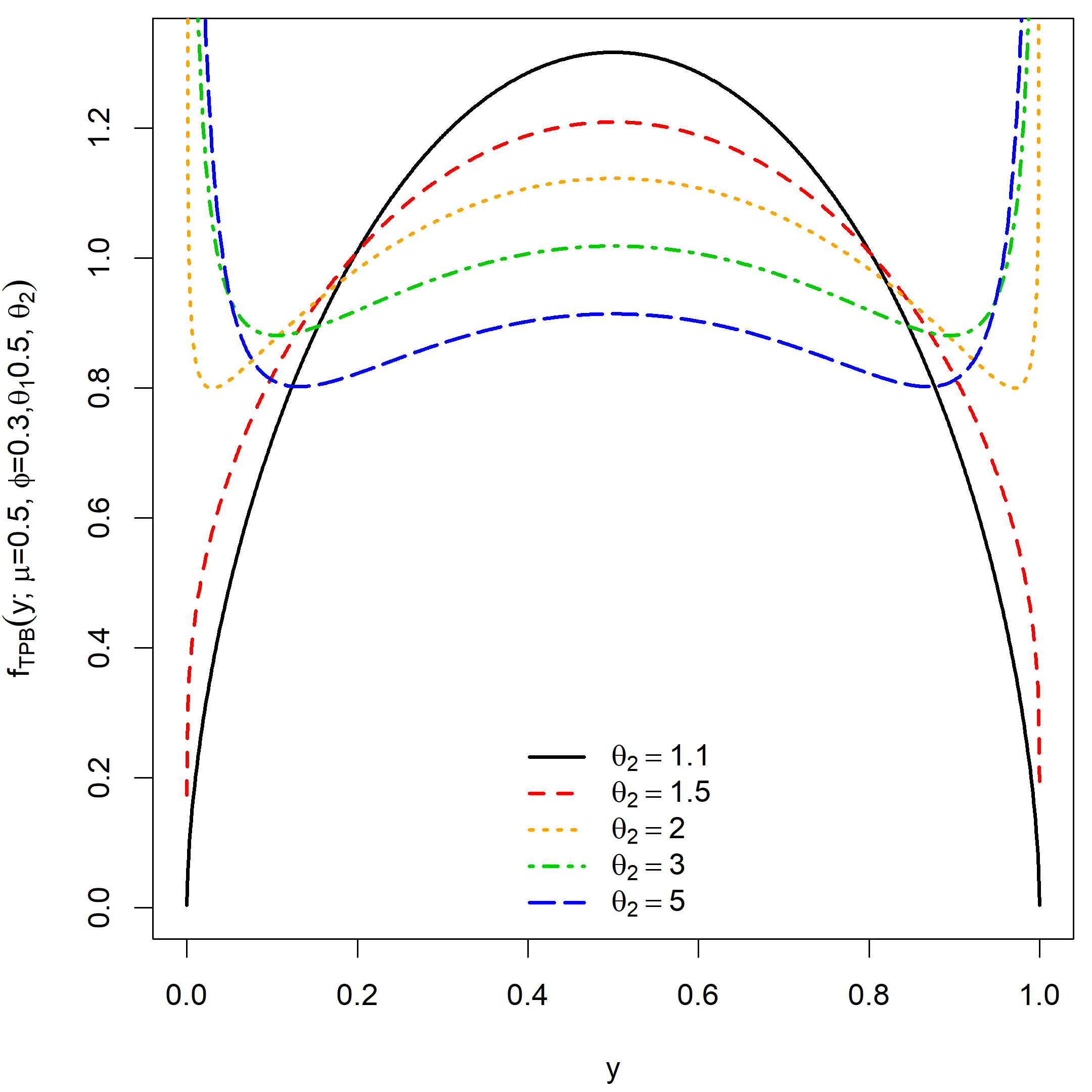}
		\caption{Varying values of $\theta_2$, when $\mu=0.5, \phi=0.3, \theta_1=0.5$. }
	\end{subfigure}
	\caption{Plots of the TPB distribution \eqref{UB-BER:PDF} for varying parameter values.}
\label{plot tpb}
\end{figure}

In this section, the mixing random variable $W$ is assumed to follow a two-point (TP) distribution. The resulting BSM is then referred to as the TPB distribution. Specifically, let $W$ be a discrete random variable defined as
\begin{equation*} \label{TBP W weight}
W=
    \begin{cases}
         \space 1 & \text{with probability $\theta_1$},\\
        \space\frac{1}{\theta_2} & \text{with probability $1- \theta_1$},
    \end{cases}
\end{equation*}
where $\theta_1 \in (0,1)$ and $\theta_2>1$. The corresponding probability mass function can be written as
\begin{equation*}\label{UB-BER: H DIST}
    h_\text{TP}(w;\boldsymbol{\theta}) = \theta_1^{\frac{w-\frac{1}{\theta_2}}{1-\frac{1}{\theta_2}}}(1-\theta_1)^{\frac{1-w}{1-\frac{1}{\theta_2}}}, \quad w\in\{1/\theta_2,1\},
\end{equation*}
where $\boldsymbol{\theta}= (\theta_1,\theta_2)$. A random variable $Y$ is then said to have a TPB distribution if its PDF is given by 
\begin{equation}\label{UB-BER:PDF}
f_{\text{TPB}}(y;\mu,\phi,\boldsymbol{\theta}) = \theta_1 f_{\text{B}}(y;\mu,\phi) +(1-\theta_1)f_{\text{B}}(y;\mu,\theta_2 \phi) , \quad \space 0< y<1,
\end{equation}
where $\space \theta_1\in(0,1)$,\space$\theta_2>1,\space \mu\in(0,1)$ and $\phi >0$. This is denoted as $Y \sim \mathcal{TPB}(\mu,\phi,\theta_1,\theta_2)$.
 The TPB distribution can be considered as a contaminated distribution, namely a two-component mixture of beta distributions, where one component represents the typical observations (the reference beta distribution) and the other, with the same mean, $\mu$, but an increased variance for the extreme observations (the contaminant distribution). For a discussion on the concept of a reference distribution, see \citet{davies1993identification}.  Here the additional parameters $\theta_1$ and $\theta_2$ have practical interpretations:
$\theta_1$ denotes the proportion of "good" points from the reference beta distribution, while $\theta_2$ denotes the degree of contamination \citep{hennig2002fixed, punzo2016parsimonious}.  Because of the assumption $\theta_2>1$, it can be viewed as an inflation parameter.  Furthermore, the TPB  distribution reduces to the $\mathcal{B}(\mu,\phi)$ when $\theta_1\to1^-$ and $\theta_2\to1^+$. The TPB distribution is also referred to as the variance-inflated beta distribution \citep{di2020robustness}. Figure \ref{plot tpb} illustrates the effects of varying  $\theta$ on \eqref{UB-BER:PDF}, while other parameters remain fixed.

An advantage of (\ref{UB-BER:PDF}) as discussed by \cite{punzo2021modeling} and \cite{mazza2017modeling}, is that given the estimates of $\mu$, $\phi$, $\theta_1$ and $\theta_2$, say $\hat{\mu}$, $\hat{\phi}$, $\hat\theta_1$ and $\hat\theta_2$, respectively, it is possible to distinguish whether a generic data-point $y$ comes from the reference beta distribution or not via the \emph{a posterioiri} probability
\begin{equation}\label{UB-BER: POST PROB}
        P(\text{$y$ comes from $\mathcal{B}(\mu,\phi)$}|\hat{\mu},\hat{\phi},\hat\theta_1,\hat\theta_2) = \frac{\hat\theta_1 f_{\text{B}}(y,\hat{\mu},\hat{\phi})}{f_{\text{TPB}}(y,\hat{\mu},\hat \phi,\hat\theta_1, \hat\theta_2)}. 
\end{equation} 
Specifically, $y$ will be considered to be from the beta reference distribution if the posterior probability (\ref{UB-BER: POST PROB}) is greater than 0.5.

\begin{figure}[h!]
	\centering
	\begin{subfigure}[h]{0.48\textwidth}
		\centering
  \includegraphics[scale=0.48]{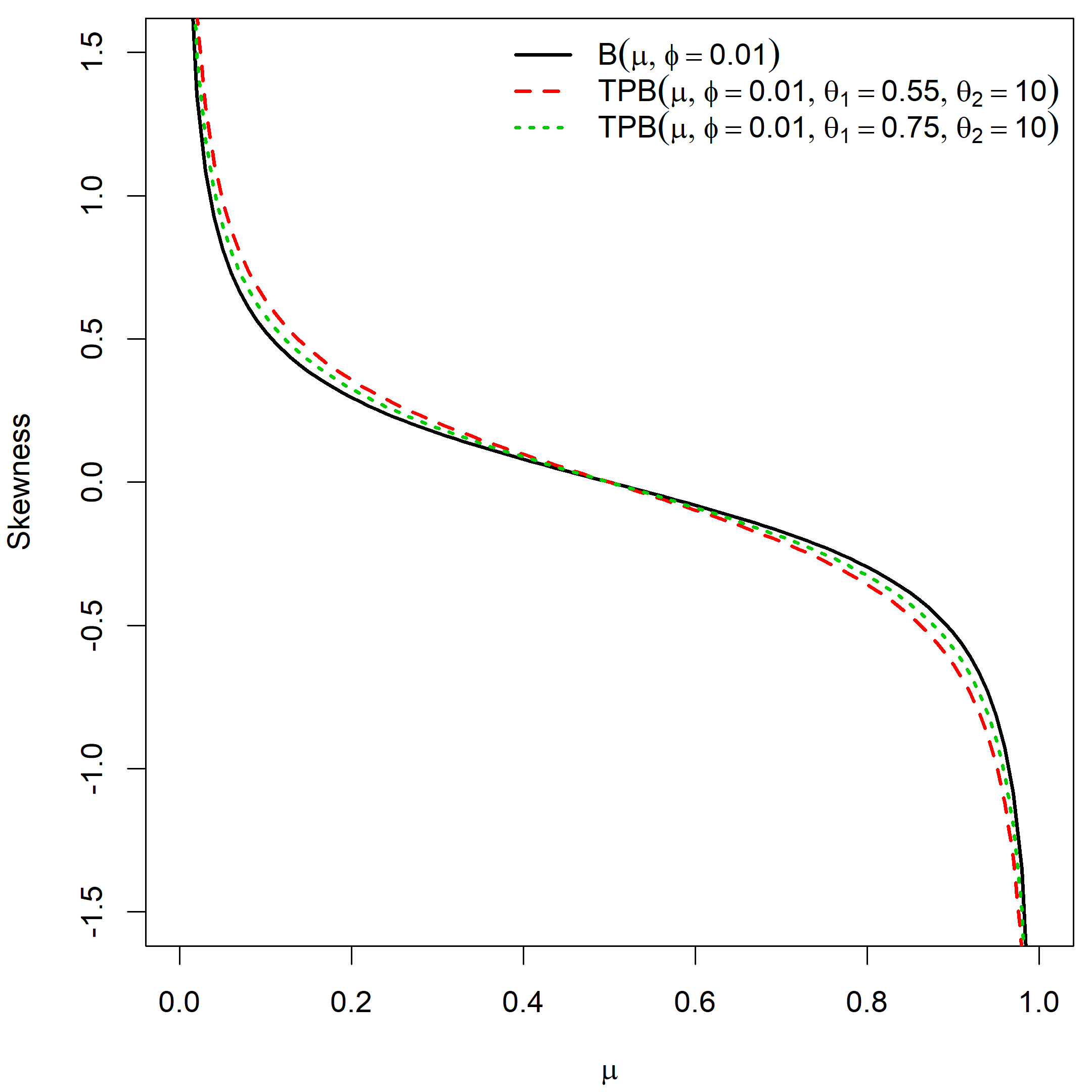}
		\caption{Skewness.  }
	\end{subfigure}
 \begin{subfigure}[h]{0.48\textwidth}
		\centering
		\includegraphics[scale=0.48]{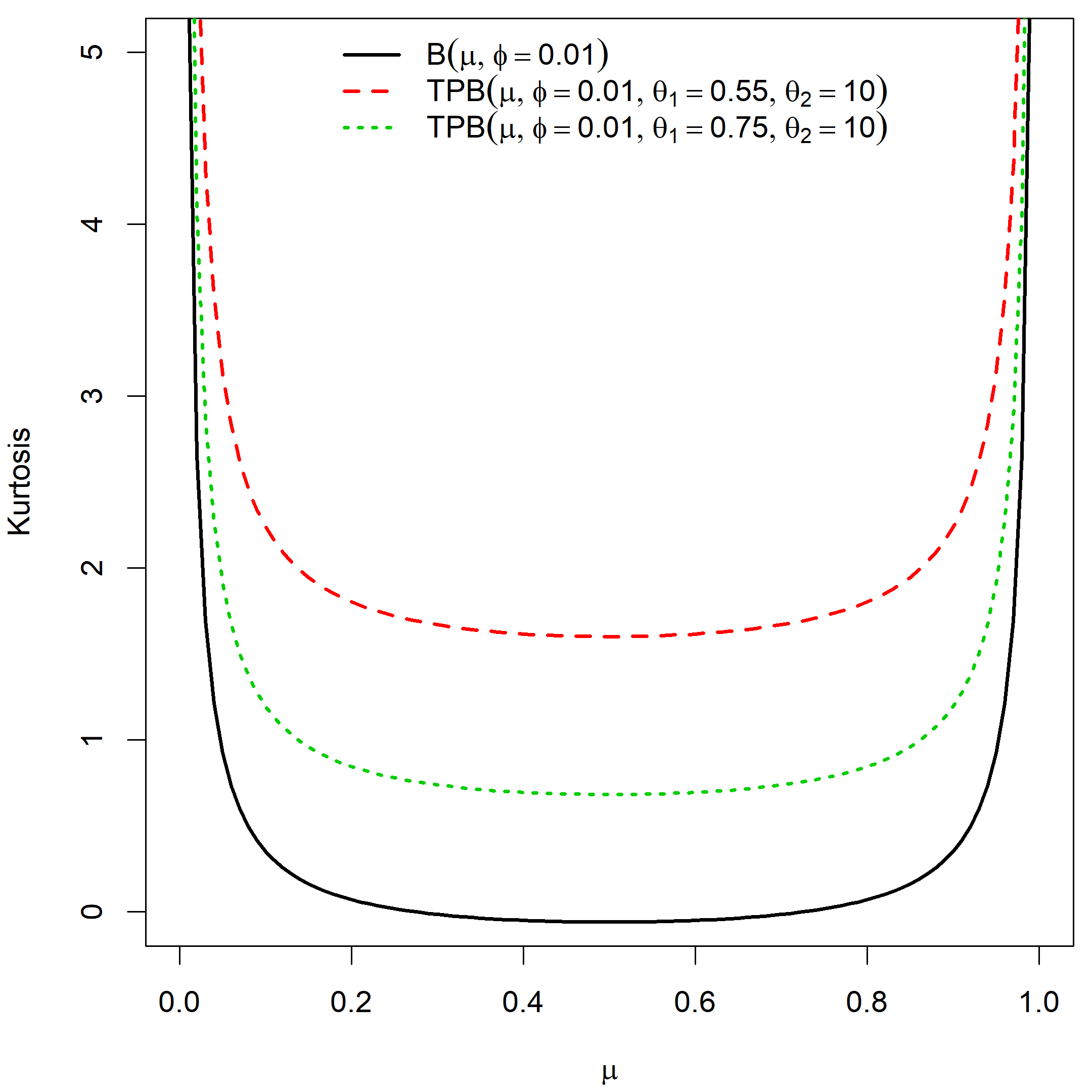}
		\caption{Kurtosis.}
	\end{subfigure}
	\caption{Examples of behaviour of Skew($X$) (on the left) and Kurt($X$) (on the right), as function of $\mu$, for fixed $\phi=0.01$ and different values of $\theta$ of the TPB distribution \eqref{UB-BER:PDF}, with $\theta_2=10$.}
\label{skew and kurt TPB1}
\end{figure}

\begin{figure}[h!]
	\centering
	\begin{subfigure}[h]{0.48\textwidth}
		\centering
  \includegraphics[scale=0.48]{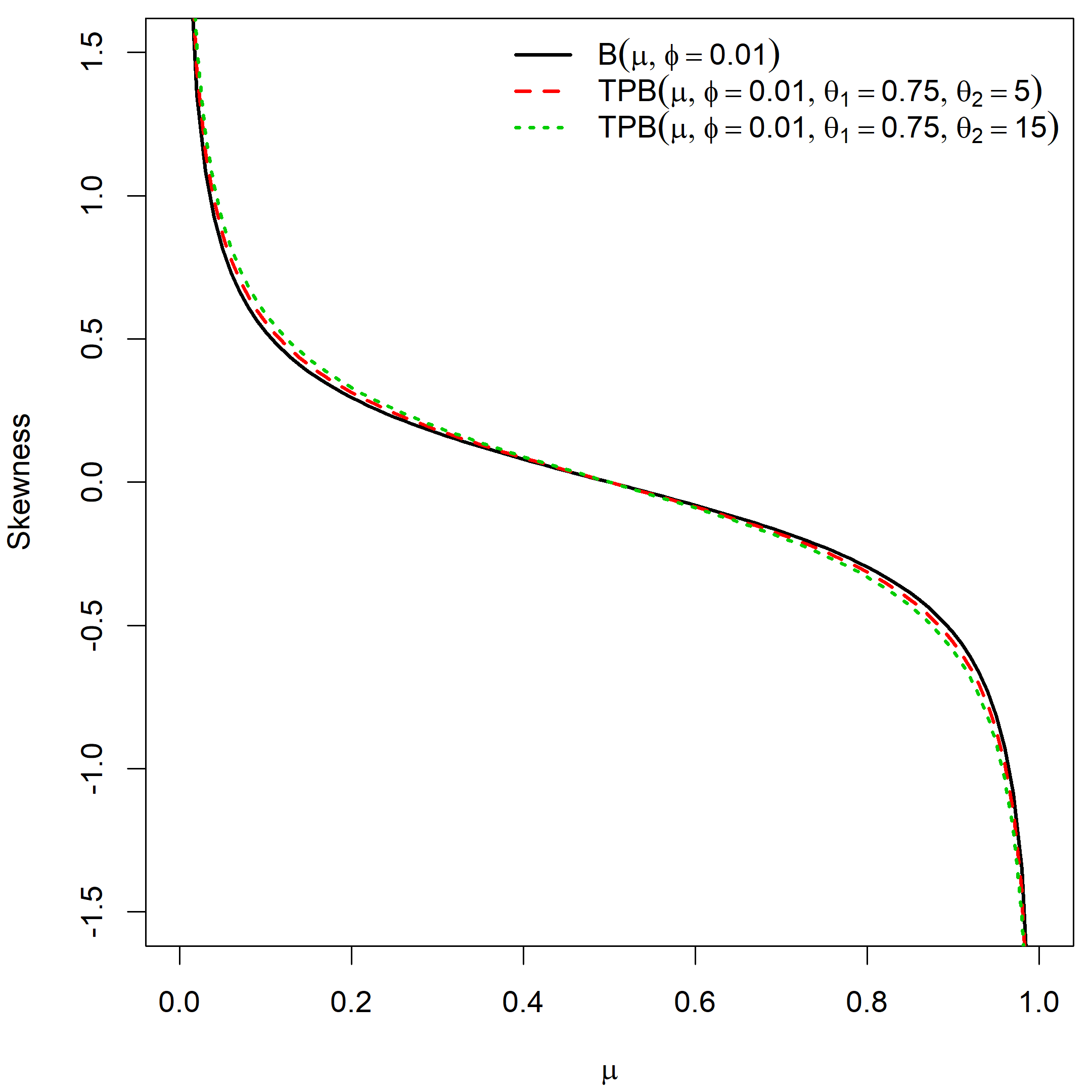}
		\caption{Skewness. }
	\end{subfigure}
 \begin{subfigure}[h]{0.48\textwidth}
		\centering
		\includegraphics[scale=0.48]{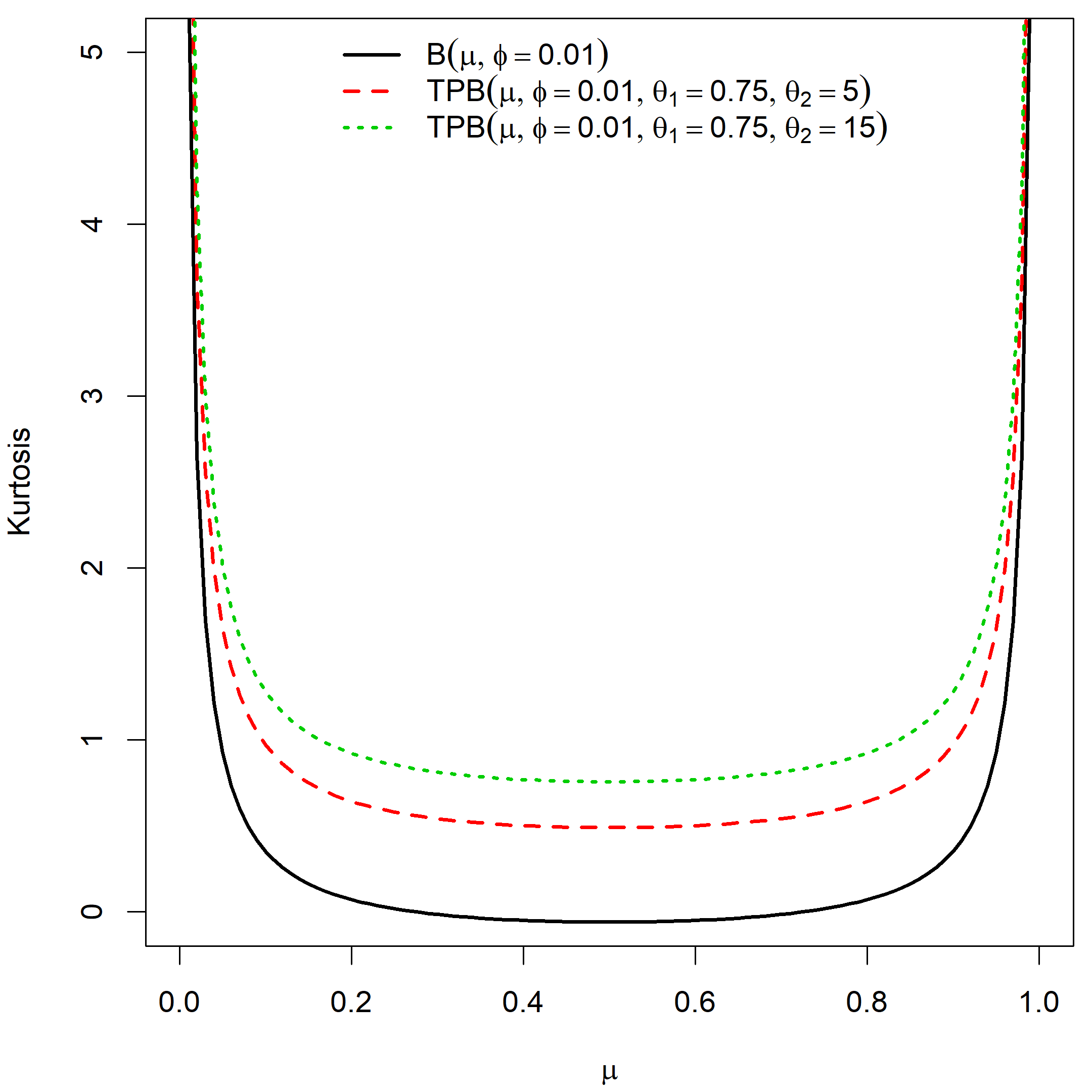}
		\caption{Kurtosis. }
	\end{subfigure}
	\caption{Examples of behaviour of Skew($X$) (on the left) and Kurt($X$) (on the right), as function of $\mu$, for fixed $\phi=0.01$ and different values of $\theta$ of the TPB distribution \eqref{UB-BER:PDF}, with $\theta_1=0.75$.}
\label{skew and kurt TPB2}
\end{figure}

As an illustration of the flexibility of the TBP distribution \eqref{UB-BER:PDF} in accommodating a larger range of skewness and kurtosis relative to that of the beta distribution, Figure \ref{skew and kurt TPB1} and \ref{skew and kurt TPB2} present examples of the skewness and kurtosis of the TBP distribution as functions of $\mu$ for varying values of $\theta_1$ and $\theta_2$, respectively. Firstly, in Figure \ref{skew and kurt TPB1}, when $\theta_1$ increases, both skewness and kurtosis move towards that of the beta model. This is expected based on \eqref{TBP W weight}, since an increase in $\theta_1$ shifts the mixing distribution towards a degenerate distribution with point mass in 1. Secondly, in Figure \ref{skew and kurt TPB2}, for fixed $\theta_1$ and increasing $\theta_2$, both skewness and kurtosis are increasing. In both cases, the skewness and kurtosis of the beta distribution serve as lower bounds, indicating that the TPB distribution offers greater skewness and kurtosis. 



\subsection{Gamma beta distribution}
\begin{figure}[h!]
    \centering
    \includegraphics[width=0.5\linewidth]{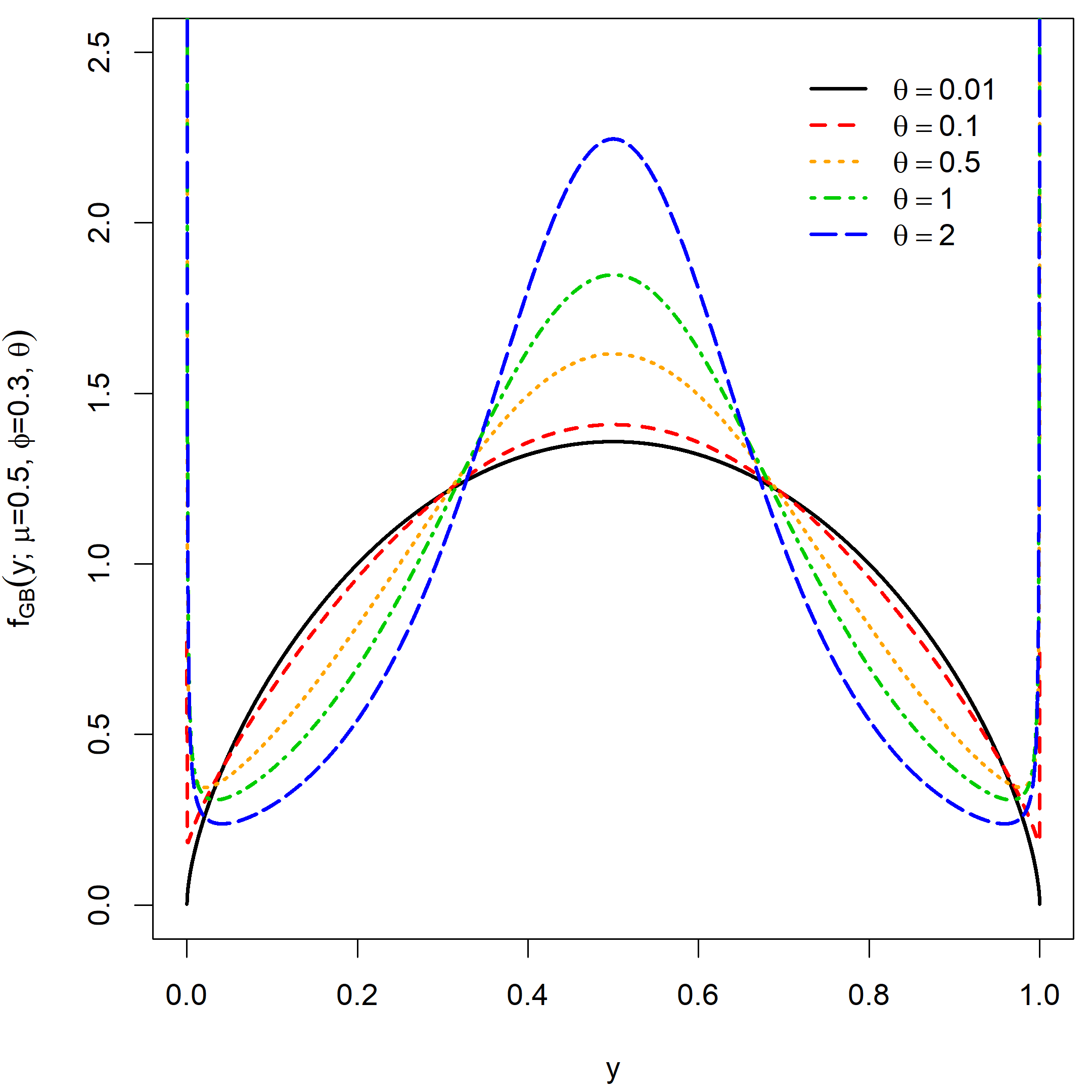}
    \caption{Plot of the GB distribution \eqref{UB-UG: PDF} for varying values of $\theta$, when $\mu=0.5$, and $\phi=0.3$}
    \label{plot gb pdf}
\end{figure}

\begin{figure}[h!]
	\centering
	\begin{subfigure}[h]{0.48\textwidth}
		\centering
  \includegraphics[scale=0.48]{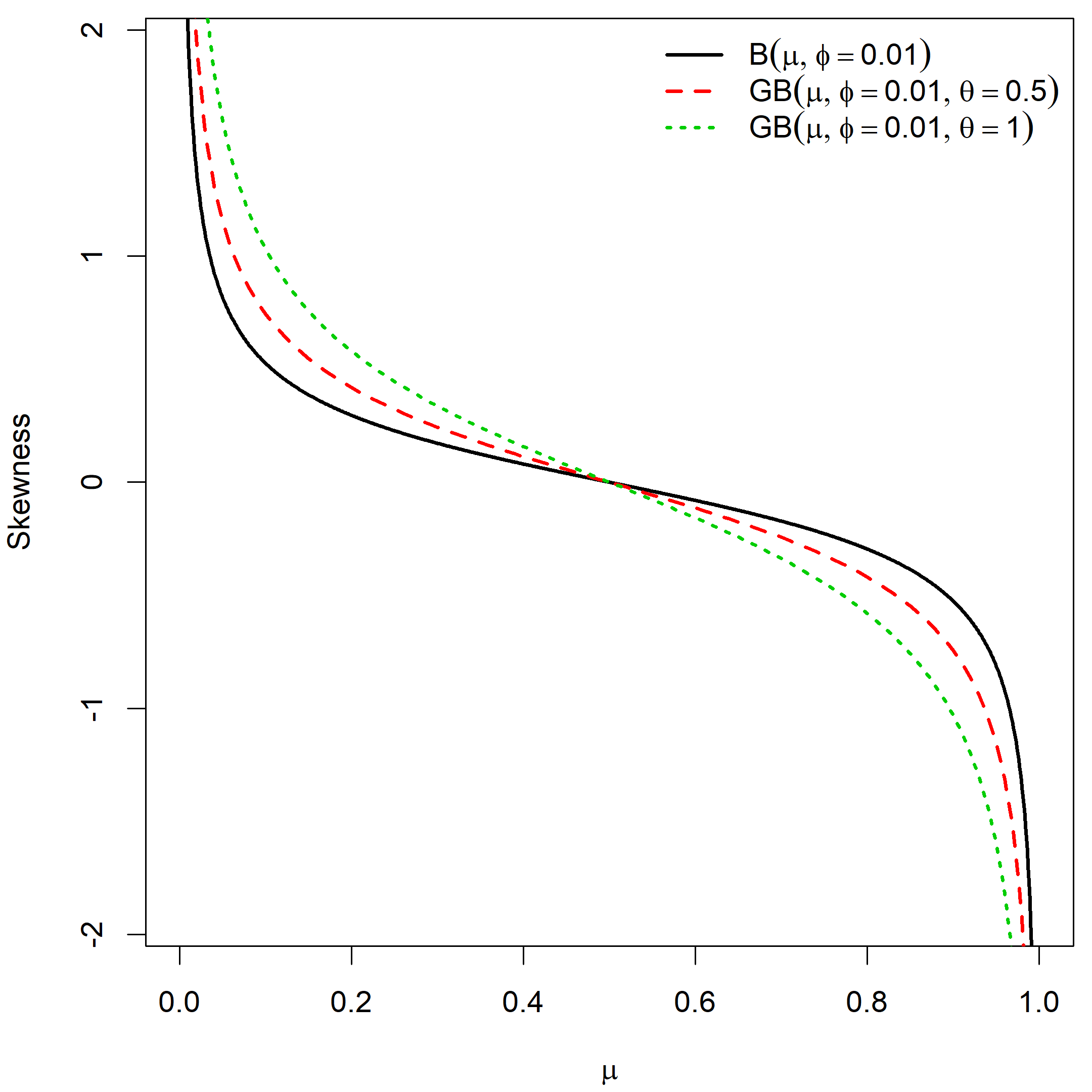}
		\caption{Skewness.}
	\end{subfigure}
 \begin{subfigure}[h]{0.48\textwidth}
		\centering
		\includegraphics[scale=0.48]{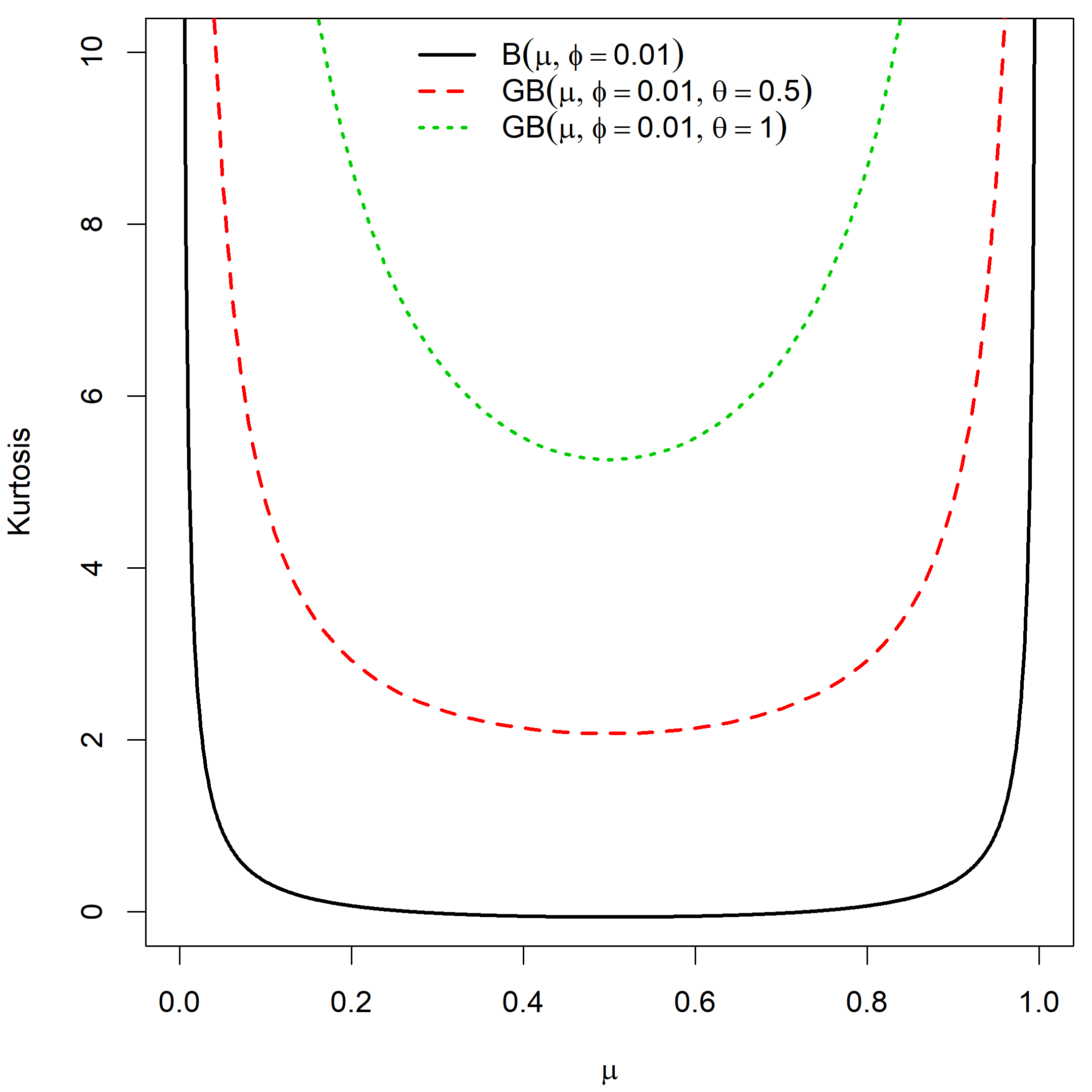}
		\caption{Kurtosis.}
	\end{subfigure}
	\caption{Examples of behaviour of Skew($X$) (on the left) and Kurt($X$) (on the right), as function of $\mu$, for fixed $\phi=0.01$ and different values of $\theta$ of the GB distribution \eqref{UB-UG: PDF}.}
\label{fig: skew kurt UB-UG}
\end{figure}

The unimodal gamma (G) distribution, a mode-based parameterisation of the gamma distribution, introduced by \cite{bagnato2013finite}, is considered as a mixing distribution in this section. The resulting distribution is then referred to as the gamma beta (GB) distribution.
Let \begin{flalign}\label{eq pdf ug}
    h_{\text{G}}(w;\theta) = \frac{w^{\frac{1}{\theta}}\exp{(\frac{-w}{\theta})}}{\theta^{\frac{1}{\theta}+1}\Gamma\left(\frac{1}{\theta}+1\right)},\quad w>0,
\end{flalign}
with $\theta>0$ be the PDF of the G distribution. If the PDF in \eqref{eq pdf ug} is considered as the mixing PDF in model \eqref{eq:2}, the PDF of the BSM becomes
\begin{equation}\label{UB-UG: PDF}
   f_{\text{GB}}(y;\mu,\phi,\boldsymbol{\theta}) =  \mathrm{E}_{\text{G}}\left[\frac{y^{\frac{\mu w}{\phi}-1}(1-y)^{\frac{w(1-\mu)}{\phi}-1}}{\mathrm{B}\left(\frac{\mu w}{\phi},\frac{w(1-\mu)}{\phi}\right)} \right], 
\end{equation}
where $\mu\in(0,1)$, $\phi >0$ and $\theta>0$. This is denoted as $X \sim \mathcal{GB}(\mu,\phi,\theta)$. Figure \ref{plot gb pdf} illustrates the effects of varying  $\theta$ on \eqref{UB-UG: PDF}, while other parameters remain fixed. As $\theta\to0^+$, the $\mathcal{GB}(\mu,\phi,\theta)$ tends to $\mathcal{B}(\mu,\phi).$
To illustrate the flexibility of the GB distribution \eqref{UB-UG: PDF} in accommodating variations in skewness and kurtosis beyond those of the standard beta model, Figure \ref{fig: skew kurt UB-UG} presents examples of the skewness and kurtosis of the GB distribution as functions of $\mu$ for varying values of $\theta$. In the case of skewness, the range of atiananle skewness values expands. When considering the kurtosis, it is observed that as $\theta \to 0^+$, the corresponding curve acts as a lower bound for the kurtosis and converges to that of the $\mathcal{B}(\mu,\phi)$ distribution.

\subsection{Log-normal beta distribution}\label{Log-normal unimodal beta distribution}

\begin{figure}[h!]
    \centering
    \includegraphics[width=0.5\linewidth]{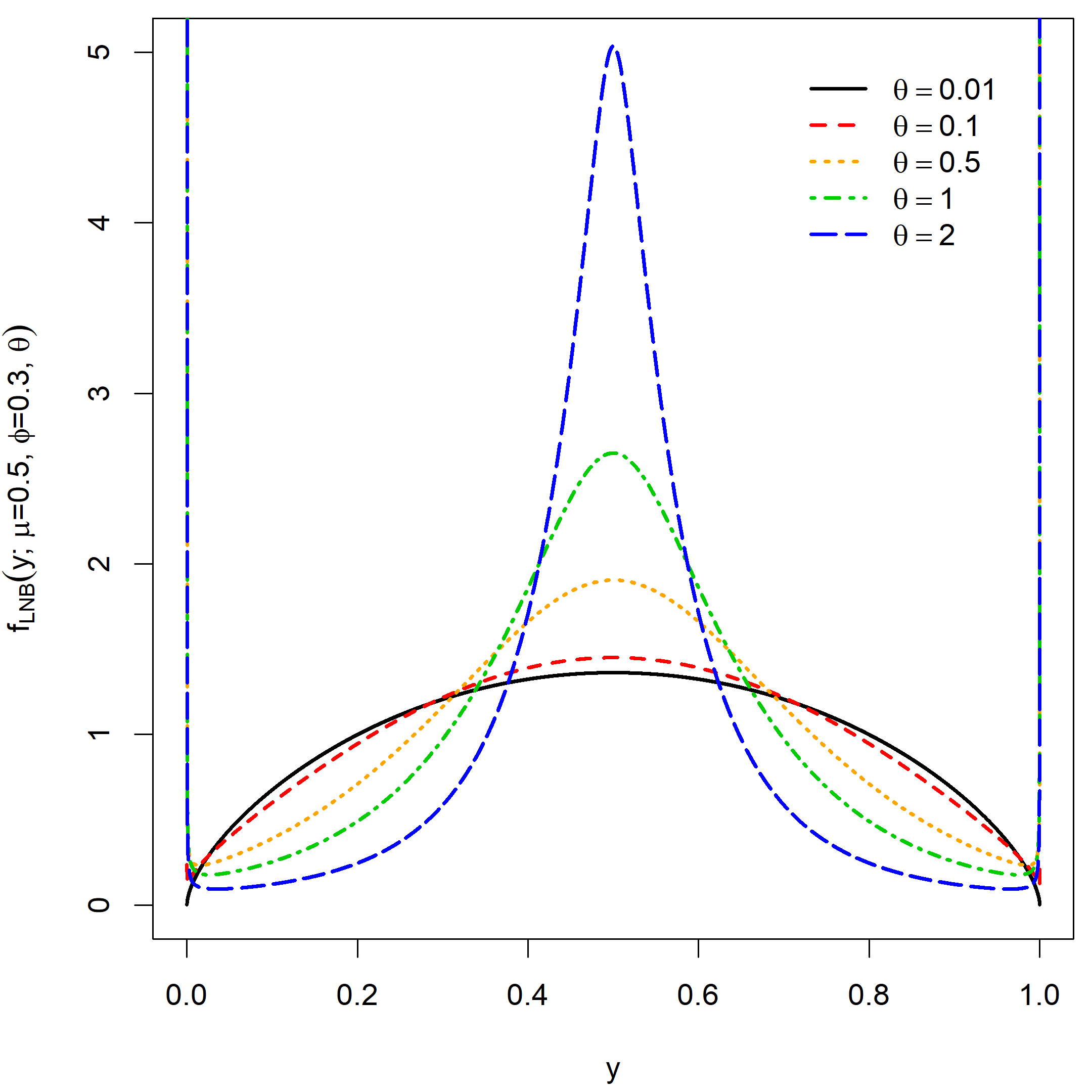}
    \caption{Plot of the LNB distribution \eqref{UB-LN: PDF} for varying values of $\theta$, when $\mu=0.5$, and $\phi=0.3$}
    \label{plot lnb pdf}
\end{figure}

 The unimodal log-normal (LN) distribution \citep{mazza2017modeling} is considered as a mixing distribution in this section. The resulting distribution is then referred to as the log-normal beta (LNB) distribution. Let
\begin{align}\label{eq pdf ln}
    h_{\text{LN}}(w;\theta) = \frac{1}{w\sqrt{2\pi\theta}}\exp{\left(-\frac{(\ln(w)-\theta)^2}{2\theta}\right)},\quad w>0,
\end{align}
with $\theta>0$, be the PDF of the LN distribution. When the PDF in \eqref{eq pdf ln} is considered as mixing density in model \eqref{eq:2}, the PDF of the BSM becomes
\begin{equation}\label{UB-LN: PDF}
   f_{\text{LNB}}(y;\mu,\phi,\boldsymbol{\theta}) =  \mathrm{E}_{\text{LN}}\left[\frac{y^{\frac{\mu w}{\phi}-1}(1-y)^{\frac{w(1-\mu)}{\phi}-1}}{\mathrm{B}\left(\frac{\mu w}{\phi},\frac{w(1-\mu)}{\phi}\right)} \right],\ \\
\end{equation}
where  $\mu\in(0,1)$, $\phi >0$, and $\theta>0$. This is denoted as $Y \sim \mathcal{LNB}(\mu,\phi,\theta)$. Figure \ref{plot lnb pdf} illustrates the effects of varying  $\theta$ on \eqref{UB-LN: PDF}, while other parameters remain fixed.  As $\theta\to0^+$, the $\mathcal{LNB}(\mu,\phi,\theta)$ tends to $\mathcal{B}(\mu,\phi).$

\begin{figure}[h!]
	\centering
	\begin{subfigure}[h]{0.48\textwidth}
		\centering
  \includegraphics[scale=0.48]{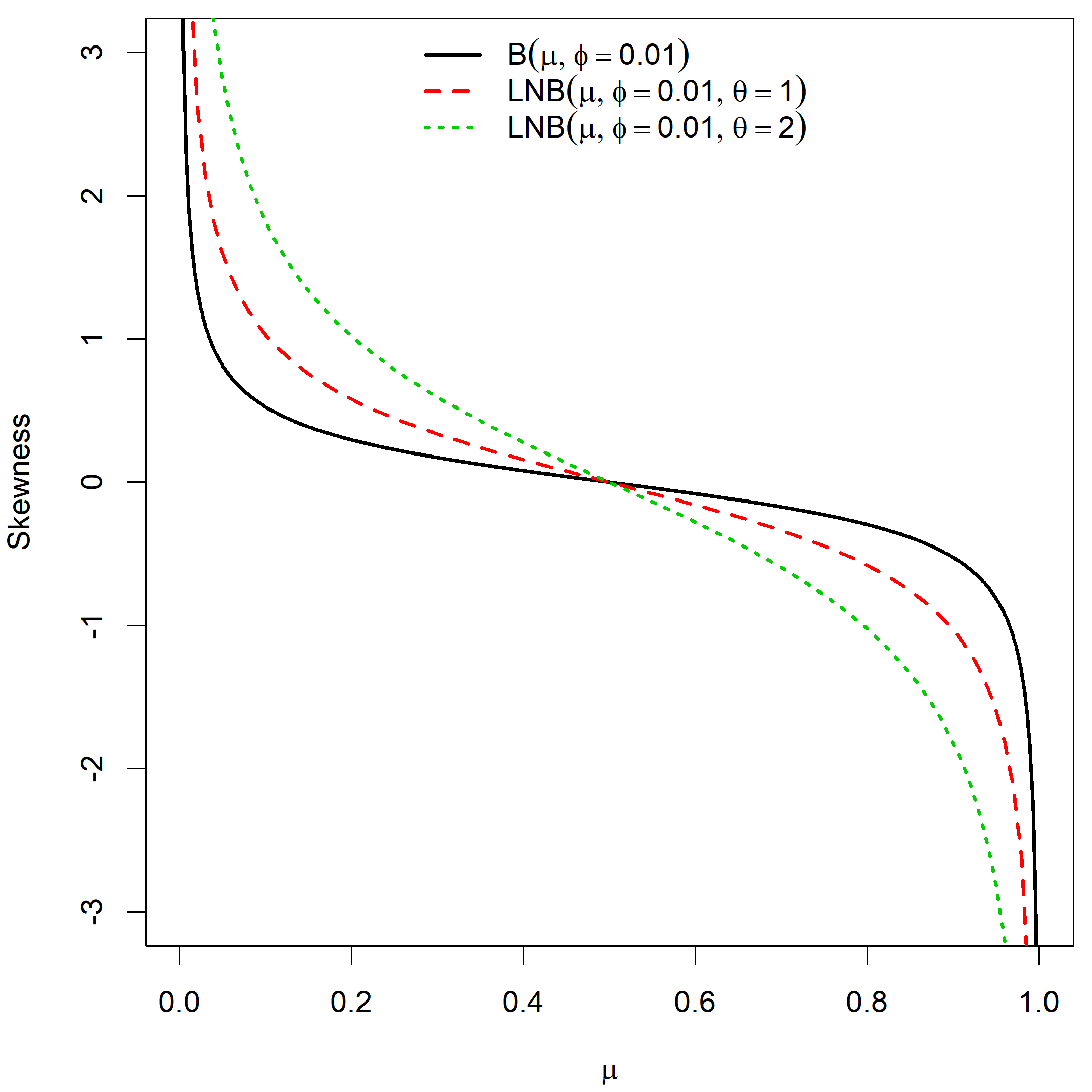}
		\caption{Skewness.  }
	\end{subfigure}
 \begin{subfigure}[h]{0.48\textwidth}
		\centering
		\includegraphics[scale=0.48]{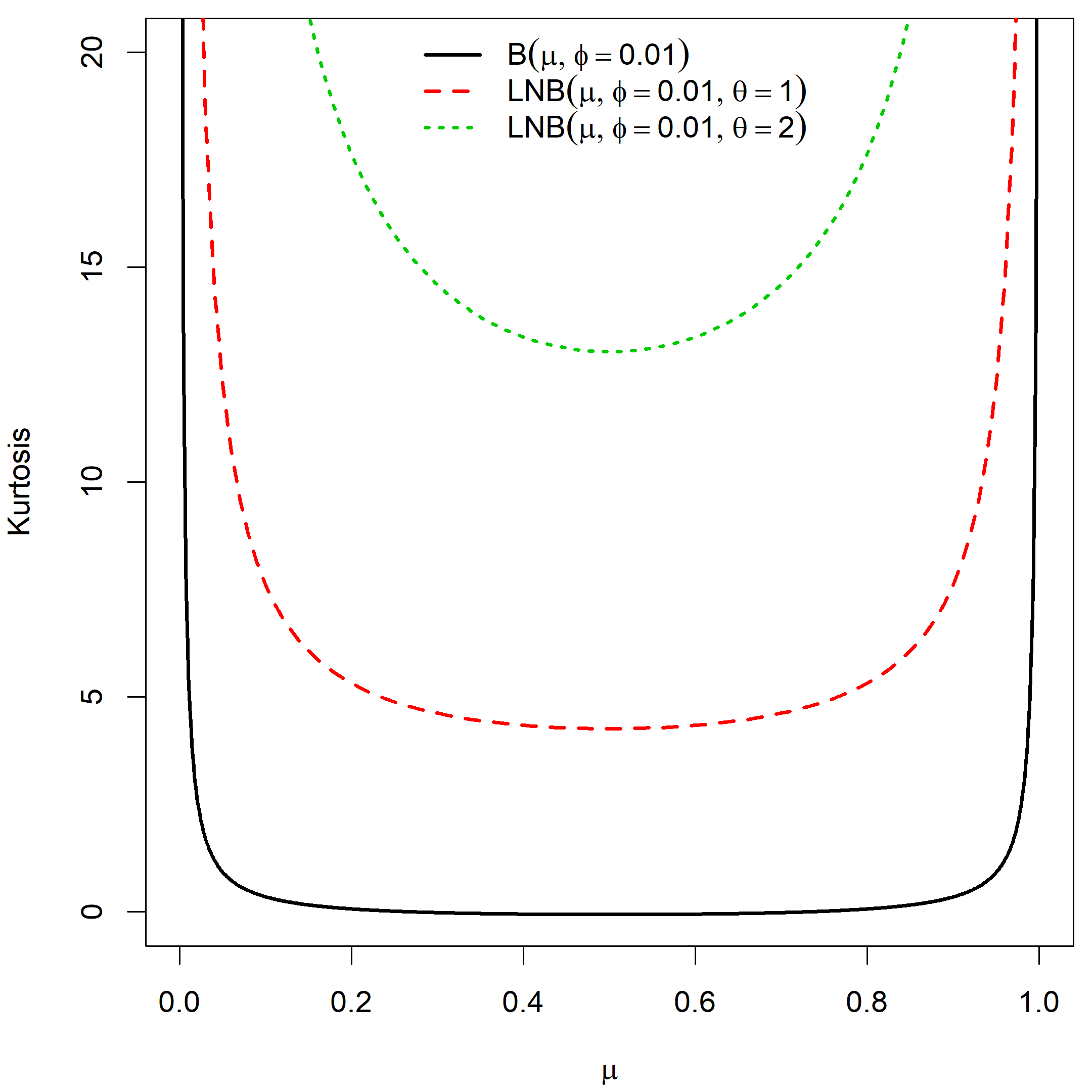}
		\caption{Kurtosis. }
	\end{subfigure}
	\caption{Examples of behaviour of Skew($X$) (on the left) and Kurt($X$) (on the right), as function of $\mu$, for fixed $\phi=0.01$ and different values of $\theta$ of the LNB distribution \eqref{UB-LN: PDF}.}
\label{fig: skew kurt UB-LN}
\end{figure}

As an illustration of the flexibility of the LNB distribution \eqref{UB-LN: PDF} in capturing a larger range of skewness and kurtosis, compared to that of the beta distribution, Figure \ref{fig: skew kurt UB-LN} presents examples of the skewness and kurtosis of the LNB distribution as functions of $\mu$ for varying values of $\theta$. 
It seems evident that the LNB distribution exhibits greater skewness and kurtosis than the beta distribution.
When considering the kurtosis, it is observed that as $\theta \to 0^+$, the corresponding curve acts as a lower bound for the kurtosis and converges to that of the $\mathcal{B}(\mu,\phi)$ distribution.


\subsection{Inverse Gaussian beta distribution}\label {Mode parameterised inverse gaussian - unimodal beta distribution}
\begin{figure}[h!]
    \centering
    \includegraphics[width=0.5\linewidth]{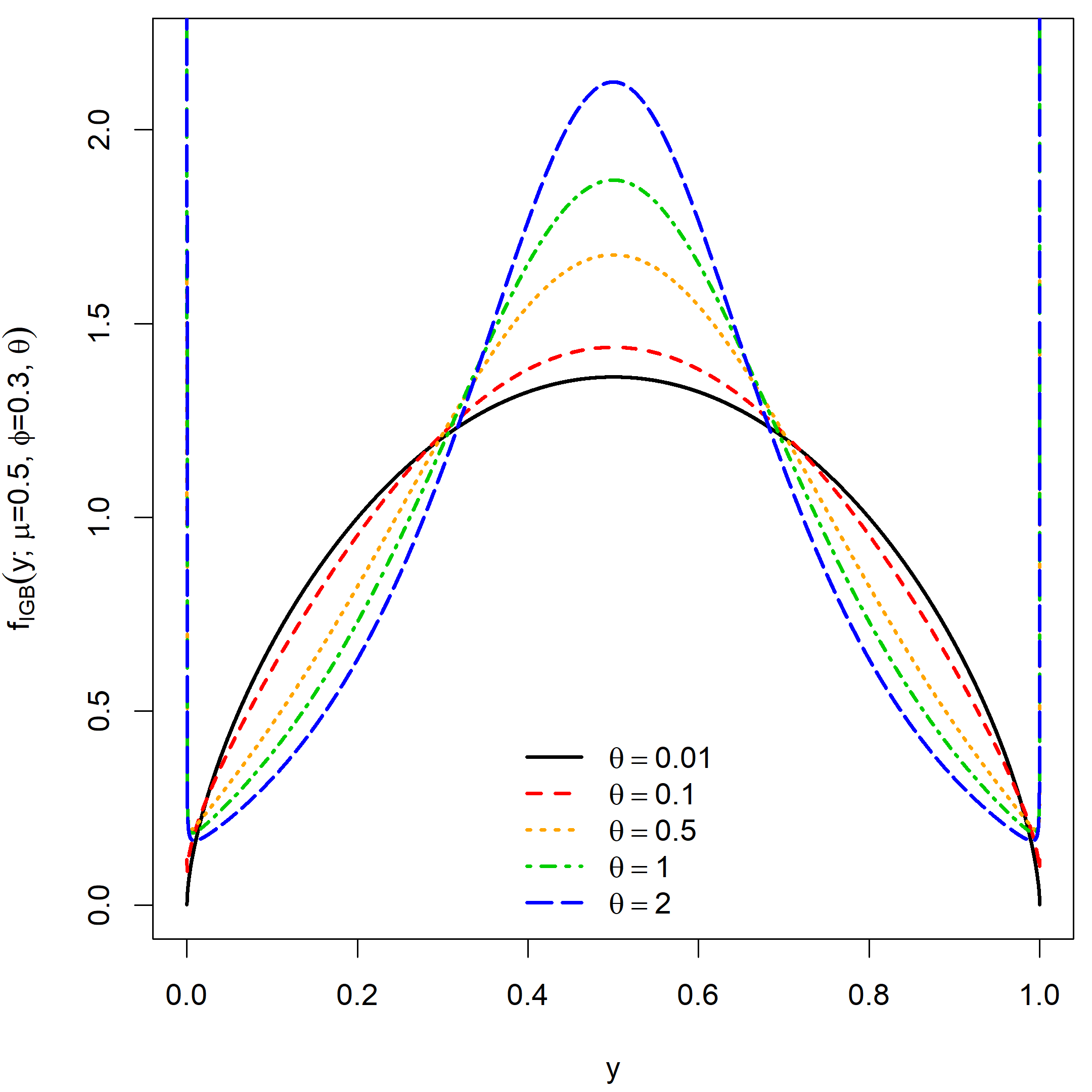}
    \caption{Plot of the IGB distribution \eqref{UB-IG: PDF} for varying values of $\theta$, when $\mu=0.5$, and $\phi=0.3$}
    \label{plot igb pdf}
\end{figure}

 The mode-parameterised inverse Gaussian (IG) distribution, a reparameterised inverse Gaussian distribution proposed by \cite{punzo2019new}, is considered as a mixing distribution in this section. The resulting distribution is then referred to as the inverse Gaussian beta (IGB) distribution. Let
\begin{flalign}\label{eq pdf uig}
    h_{\text{IG}}(w;\theta) = \sqrt{\frac{3\theta+1}{2\pi\theta w^{3}}}\exp{\left(-\frac{\left(w-\sqrt{3\theta+1}\right)^{2}}{2\theta w}\right)},\quad w>0,
\end{flalign}
with $\theta>0$, be the PDF of the IG distribution. When the PDF in \eqref{eq pdf uig} is considered as the mixing distribution in model\eqref{eq:2}, the PDF of the BSM becomes
\begin{equation}\label{UB-IG: PDF}
   f_{\text{IGB}}(y;\mu,\phi,\theta) =  \mathrm{E}_{\text{IG}}\left[\frac{y^{\frac{\mu w}{\phi}-1}(1-y)^{\frac{w(1-\mu)}{\phi}-1}}{\mathrm{B}\left(\frac{\mu w}{\phi},\frac{w(1-\mu)}{\phi}\right)}\right],\\
\end{equation}
 where $\mu\in(0,1)$, $\phi >0$ and $\theta>0$. This is denoted as $Y \sim \mathcal{{IGB}}(\mu,\phi,\theta)$. Figure \ref{plot igb pdf} illustrates the effects of varying  $\theta$ on \eqref{UB-IG: PDF}, while other parameters remain fixed. 

\begin{figure}[h!]
	\centering
	\begin{subfigure}[h]{0.48\textwidth}
		\centering
  \includegraphics[scale=0.48]{ 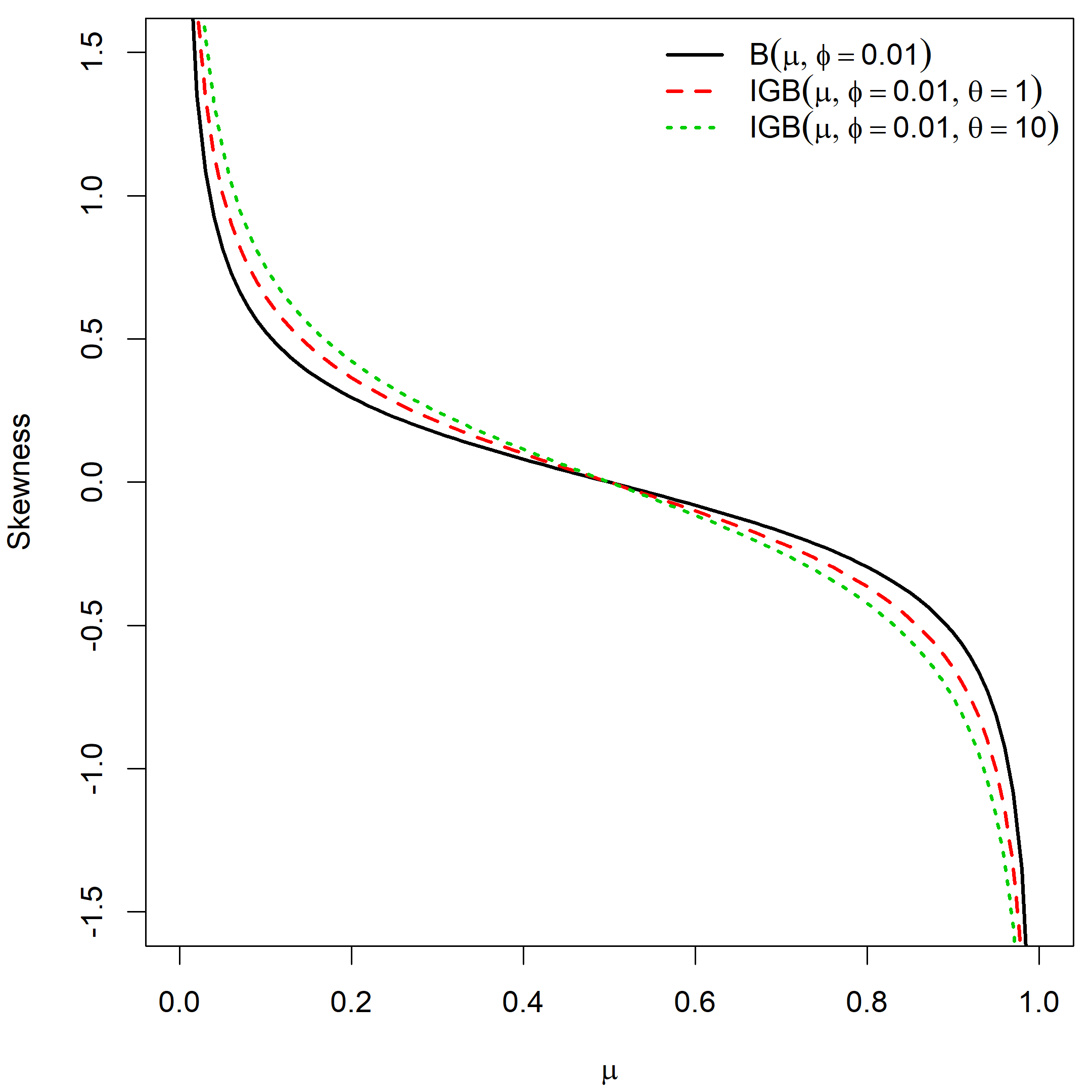}
		\caption{Skewness.  }
	\end{subfigure}
 \begin{subfigure}[h]{0.48\textwidth}
		\centering
		\includegraphics[scale=0.48]{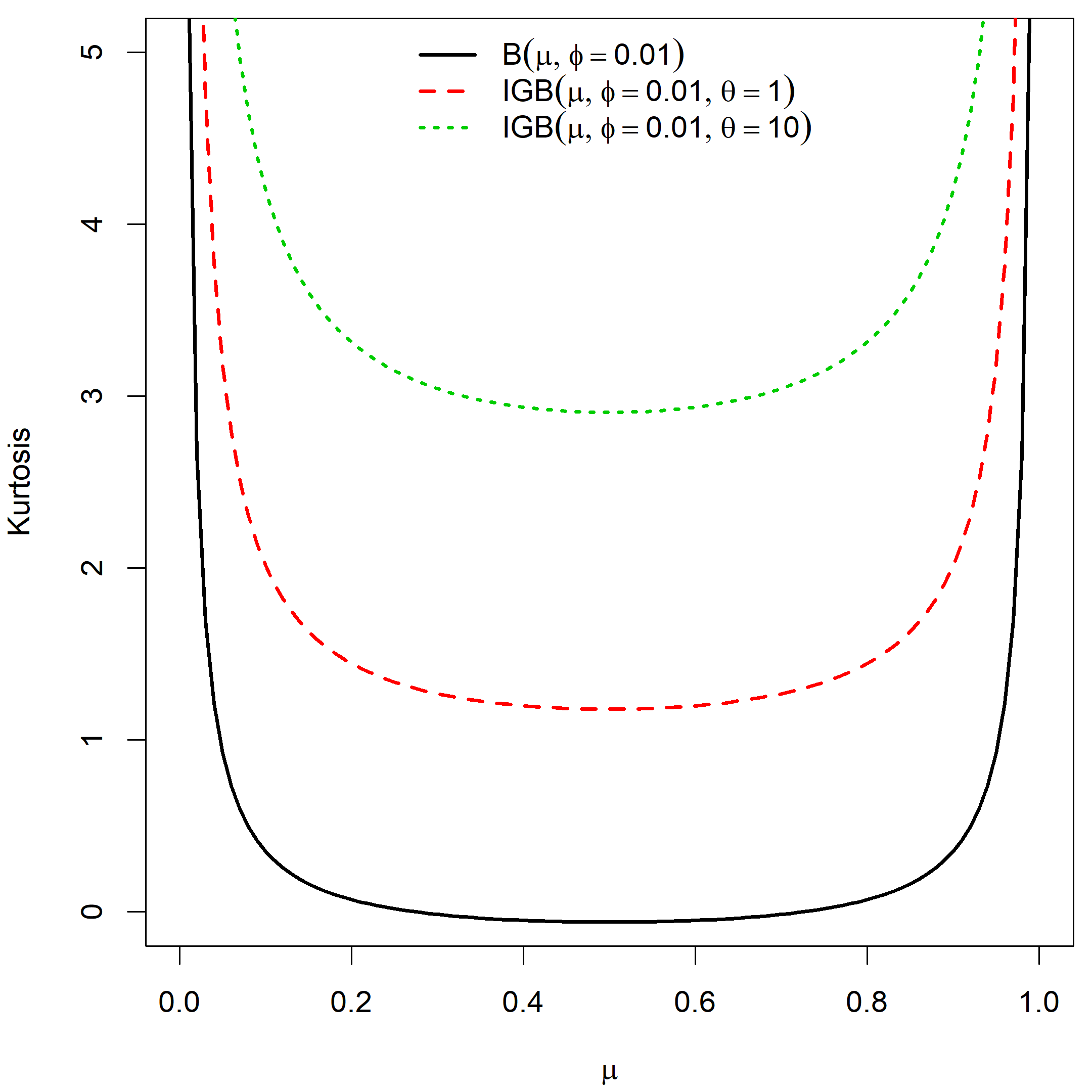}
		\caption{Kurtosis. }
	\end{subfigure}
	\caption{Examples of behaviour of Skew($X$) (on the left) and Kurt($X$) (on the right), as function of $\mu$, for fixed $\phi=0.01$ and different values of $\theta$ of the IGB distribution \eqref{UB-IG: PDF}.}
\label{fig: skew kurt UB-IGB}
\end{figure}

As an illustration of the flexibility of the IGB distribution \eqref{UB-IG: PDF} in capturing a larger range of skewness and kurtosis, compared to that of the beta model, Figure \ref{fig: skew kurt UB-IGB} presents examples of the skewness and kurtosis of the IGB distribution as functions of $\mu$ for varying values of $\theta$. In Figure \ref{fig: skew kurt UB-IGB}, when $\theta$ increases, the range of attainable skewness values for the IGB distribution expands. As in previous cases, skewness changes sign on either side of $\mu=0.5$. When considering the kurtosis specifically, it is observed that as $\theta \to 0^+$, the corresponding curve acts as a lower bound for the kurtosis and converges to that of the $\mathcal{B}(\mu,\phi)$ distribution. 


\section{Maximum likelihood estimation}\label{sec:Maximum likelihood estimation}
In this section, we outline the framework for parameter estimation via maximum likelihood estimation (MLE) for the more general BSM regression model. Specifically, we present a framework for obtaining estimates through the use of direct numerical maximization for the GB, LNB, and IGB models, while an EM algorithm for the TPB regression model is outlined in Section~\ref{sec:EM UBSM}.
\subsection{Direct numerical maximization}
Let $(\boldsymbol{x}_1',y_1),\dots,(\boldsymbol{x}_n',y_n)$ be an observed sample from the BSM regression model. The likelihood function is then
\begin{equation*}\label{eq:lik_ubsm}
L(\boldsymbol{\beta},\phi,\boldsymbol{\theta}) = 
\prod_{i=1}^n
f_{\text{BSM}}(y_i;\mu(\boldsymbol{x}_i;\boldsymbol\beta),\phi,\boldsymbol{\theta}).
\end{equation*}
Taking logarithms, the corresponding log-likelihood function is
\begin{equation*}\label{eq:loglik_ubsm}
\ell(\boldsymbol{\beta},\phi,\boldsymbol{\theta})=
\sum_{i=1}^n
\log \left[f_{\text{BSM}}(y_i;\mu(\boldsymbol{x}_i;\boldsymbol\beta),\phi,\boldsymbol{\theta})\right].
\end{equation*}

The computational ease and availability of optimization routines in \texttt{R} software make direct numerical maximization a practical choice to obtain parameter estimates. While estimates for the GB, LNB, and IGB distributions can also be obtained by making use of the EM algorithm, we opted to make use of direct numerical maximization instead by using the \texttt{optim()} function included in the \textbf{stats} package in \texttt{R}. 
An EM algorithm for the TPB distribution, however, follows as an illustration.
\subsection{An EM Algorithm for the TPB distribution}\label{sec:EM UBSM}
For the application of the EM algorithm, it is convenient to view the observed data as incomplete. In this case, the source of incompleteness stems from the fact that we do not know if the generic data point $(\boldsymbol{x}_i',y_i)$ comes from the reference beta component or not. To denote this source of incompleteness, we use an indicator variable $\boldsymbol z=(z_1,\dots, z_n)$ so that $z_i=1$ if $(\boldsymbol{x}_i',y_i)$ comes from the reference beta regression model and $z_i=0$ otherwise. The complete-data are then given by  $(\boldsymbol{x}_1',y_1,z_1),\dots,(\boldsymbol{x}_n',y_n,z_n)$. It follows that the complete-data likelihood function can then be written as
\begin{equation*}
L_{c}(\beta,\phi,\theta_1,\theta_2) = \prod_{i=1}^{n} [\theta_1 f_{\text{B}}(y_{i};\mu(\boldsymbol{x}_i;\boldsymbol{\beta}),\phi)]^{z_{i}} [(1-\theta_1)f_{\text{B}}(y_{i};\mu(\boldsymbol{x}_i;\boldsymbol{\beta}),\theta_2 \phi)]^{1-z_{i}}.
\end{equation*}
The complete log-likelihood then follows as:
\begin{align*}\label{eq loglike ub-ber}
\ell_c(\boldsymbol{\beta},\phi,\theta_1,\theta_2)=\ell_{c_1}(\theta_1)+\ell_{c_2}(\boldsymbol{\beta},\phi,\theta_2)
\end{align*}
where \begin{flalign*}
\ell_{c_1}(\theta_1) = \sum_{i=1}^{n}[z_{i}\mathrm{log}(\theta_1) + (1-z_{i})\mathrm{log}(1-\theta_1)]
\end{flalign*}
and
\begin{flalign*}
    \ell_{c_2}(\boldsymbol{\beta},\phi,\theta_2) = \sum_{i=1}^{n}  [z_{i}\mathrm{log}f_{\text{B}}(y_{i};\mu(\boldsymbol{x}_1;\boldsymbol{\beta}),\phi) + (1-z_{i})\mathrm{log}f_{\text{B}}(y_{i};\mu(\boldsymbol{x}_i;\boldsymbol{\beta}),\theta_2 \phi)].
\end{flalign*} 
The algorithm iterates between the E-step and M-step until convergence. The steps for the $(r+1)^{\text{th}}$ iteration of the algorithm are detailed below.
\subsubsection*{E-step}
In the E-step, the conditional expectation of the complete-data log-likelihood function is computed as
\begin{align}
Q\left(\boldsymbol{\beta},\phi,\theta_1,\theta_2|\boldsymbol{\beta}^{(r)}, \phi^{(r)}, \theta_1^{(r)}, \theta_2^{(r)}\right)=Q_1\left(\theta_1|\boldsymbol{\beta}^{(r)}, \phi^{(r)}, \theta_1^{(r)}, \theta_2^{(r)}\right)+Q_2\left(\boldsymbol{\beta},\phi,\theta_2|\boldsymbol{\beta}^{(r)}, \phi^{(r)}, \theta_1^{(r)}, \theta_2^{(r)}\right)
\end{align}
for the $(r+1)^{\text{th}}$ iteration. $Q\left(\boldsymbol{\beta},\phi,\theta_1,\theta_2|\boldsymbol{\beta}^{(r)}, \phi^{(r)}, \theta_1^{(r)}, \theta_2^{(r)}\right)$ is obtained by sustituting $z_i$ in \eqref{eq loglike ub-ber} by the expected \textit{a posteriori} probability for a point to come from the reference beta component
\begin{flalign*}
    \mathrm{E}\left(Z_{i}|y_i,\boldsymbol{x}_i,\boldsymbol{\beta}^{(r)}, \phi^{(r)},\theta_1^{(r)}, \theta_2^{(r)}\right)&=\frac{\theta_1^{(r)}f_{\text{B}}\left(y_{i};\mu(\boldsymbol{x}_i;\boldsymbol{\beta}^{(r)}),\phi^{(r)}\right)}{\theta_1^{(r)} f_{\text{B}}\left(y_{i};\mu(\boldsymbol{x}_i;\boldsymbol{\beta}^{(r)}),\phi^{(r)}\right) +\left(1-\theta_1^{(r)}\right)f_{\text{B}}\left(y_{i};\mu(\boldsymbol{x}_i,\boldsymbol{\beta}^{(r)}),\theta_2^{(r)}\phi^{(r)}\right)}\\
    & = : z_{i}^{(r)}.
\end{flalign*}
\subsubsection*{M-step}
An update $\theta_1^{(r+1)}$ for $\theta_1$ is calculated by independently maximizing
  \begin{align*}
  Q_{1}\left(\theta_1|\boldsymbol{\beta}^{(r)}, \phi^{(r)}, \theta_1^{(r)}, \theta_2^{(r)}\right)=\sum_{i=1}^{n}\left[z_{i}^{(r)}\mathrm{log}\left(\theta_1^{(r)}\right) + \left(1-z_{i}^{(r)}\right)\mathrm{log}\left(1-\theta_1^{(r)}\right)\right]
  \end{align*}
  with respect to $\theta_1$ and subject to the constraints on this parameter. It follows that an update for the $(r+1)^{\text{th}}$ iteration is given as
  \begin{align*}
      \theta_1^{(r+1)} &=\frac{\sum_{i=1}^{n} z_{i}^{(r)}}{n},
  \end{align*}
Similarly, updates $\boldsymbol{\beta}^{(r+1)}, \phi^{(r+1)}$, and $\theta_2^{(r+1)}$ are obtained by maximising $Q_{2}\left(\boldsymbol{\beta},\phi,\theta_2|\boldsymbol{\beta}^{(r)}, \phi^{(r)}, \theta_1^{(r)}, \theta_2^{(r)}\right)$ directly. 
\section{Simulation study: a sensitivity analysis}\label{sec: Sensitivity analysis} 
In this study, we perform a sensitivity analysis to assess how atypical values affect the estimates of the beta and BSM regression models. We generate 500 datasets of size $n = 500$ from the beta regression model with an intercept of $\beta_0 = 0.5$, and a continuous covariate generated by a standard normal distribution, with slope $\beta_1=1$. A proportion of the generated data is then randomly replaced according to one of the following scenarios:\\ 
 \begin{enumerate}
     \item 1\% of the generated $Y$-values are randomly substituted by data generated from a uniform distribution over the interval $(0,1).$\\
\item 5\% of the generated $Y$-values are randomly substituted by data generated from a uniform distribution over the interval $(0,1).$\\
 \end{enumerate}
 
\begin{figure}[h!]
	\centering
	\begin{subfigure}[h]{0.48\textwidth}
		\centering
  \includegraphics[scale=0.48]{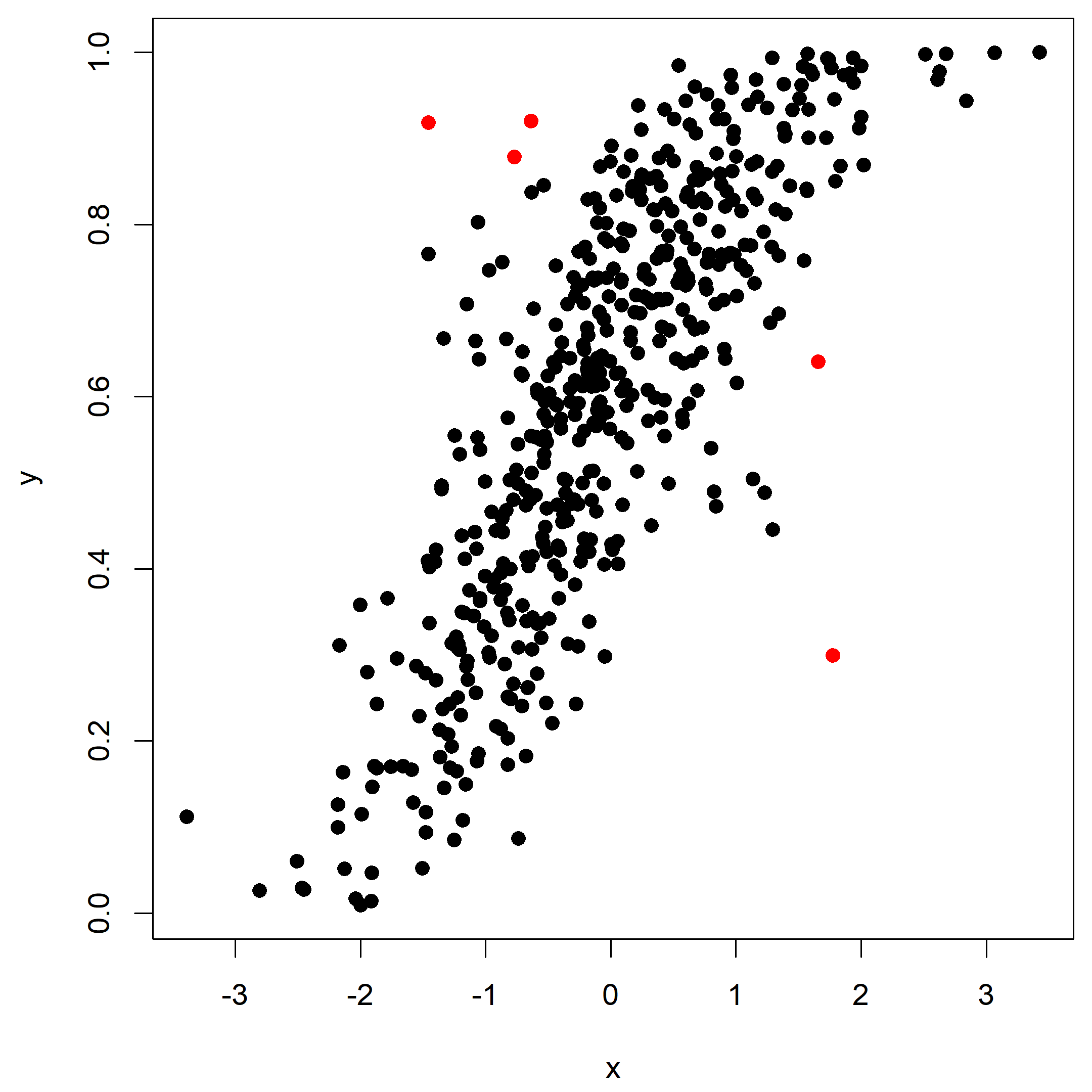}
		\caption{1\%.  }
	\end{subfigure}
 \begin{subfigure}[h]{0.48\textwidth}
		\centering
		\includegraphics[scale=0.48]{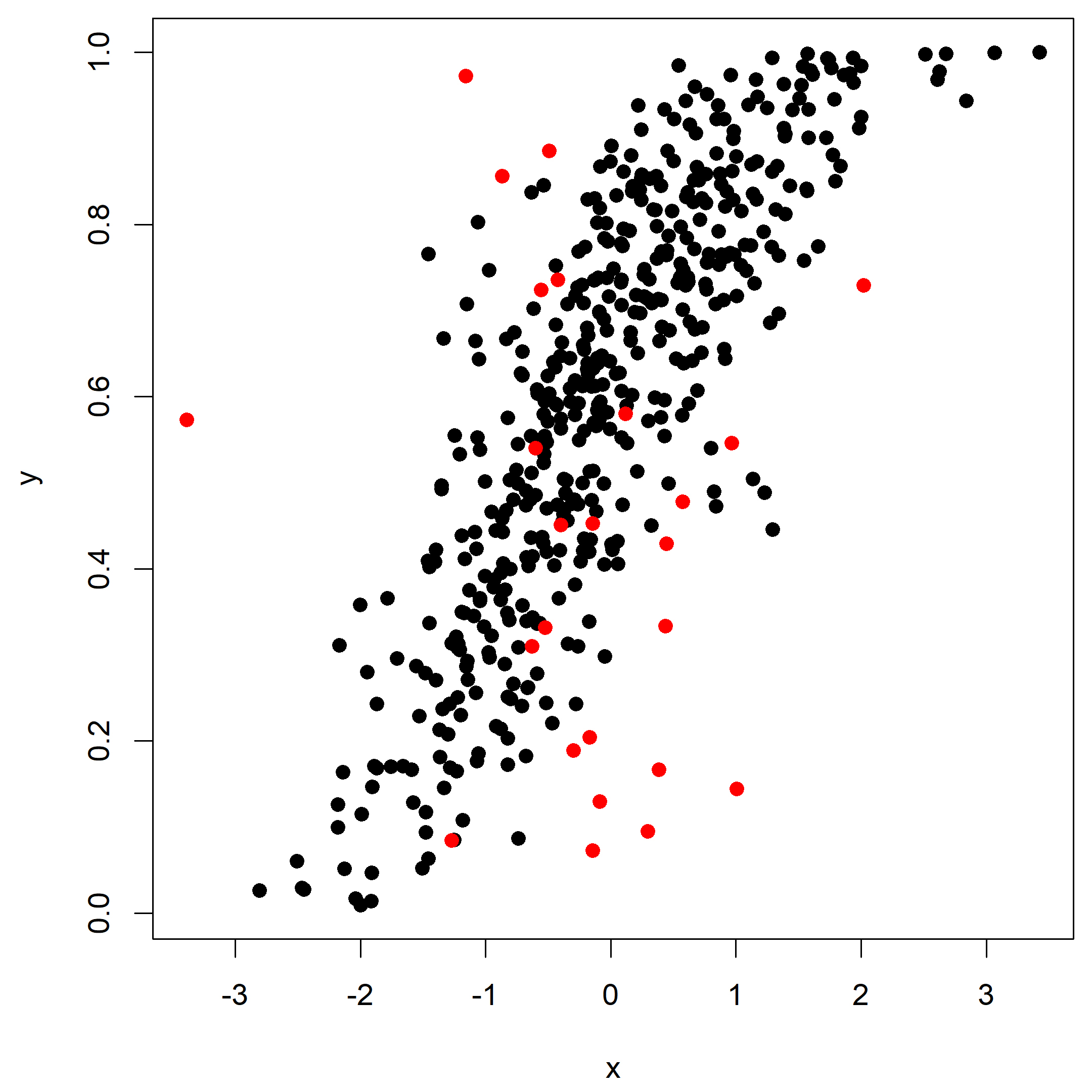}
		\caption{5\%. }
	\end{subfigure}
	\caption{Example of simulated beta data with a different percentage of artificially added outliers (in red).}
\label{plot sensitivity analysis}
\end{figure}

An example of the aforementioned scheme is illustrated in \figurename~\ref{plot sensitivity analysis} where the substituted observations are highlighted in red. Tables \ref{tab: sens analysis one percent} and \ref{tab: sens analysis five percent} report the results of the sensitivity analysis, assessed using bias and mean squared error (MSE). The results indicate that the presence of the atypical values adversely affects the estimates of the beta regression model, with the impact becoming more pronounced as the percentage of anomalous values increases from 1\% to 5\%. In contrast, the BSM regression models consistently exhibit lower bias and MSE, indicating that their greater flexibility makes them more reliable when handling data contaminated with atypical values.


\begin{table}[H]
\centering
\caption{Bias and mean squared error (MSE) of the regression parameter estimates for the beta and BSM models when 1\% of the response values are replaced by atypical observations.}
\begin{tabular}{llrrrrr}
\hline
\multicolumn{1}{l}{}  &       & \multicolumn{5}{c}{1\%}                                                                                                      \\ \cline{3-7} 
\multicolumn{1}{l}{}  &       & \multicolumn{1}{c}{B} & \multicolumn{1}{c}{TPB} & \multicolumn{1}{c}{GB} & \multicolumn{1}{c}{LNB} & \multicolumn{1}{c}{IGB} \\ \hline
\multirow{3}{*}{bias} & $\hat\beta_0$ & -0.0083               & -0.0026                 & -0.0039                & -0.0046                 & -0.0047                 \\
                      & $\hat\beta_1$ & -0.0170               & -0.0066                 & -0.0089                & -0.0097                 & -0.0101                 \\
                      & $\hat\phi$   & 0.0969                & 0.0859                  & 0.0949                 & 0.1038                  & 0.1036                  \\
\multicolumn{1}{l}{}  &       & \multicolumn{1}{l}{}  & \multicolumn{1}{l}{}    & \multicolumn{1}{l}{}   & \multicolumn{1}{l}{}    & \multicolumn{1}{l}{}    \\
\multirow{3}{*}{MSE}  & $\hat\beta_0$ & 0.0011                & 0.0009                  & 0.0010                 & 0.0010                  & 0.0010                  \\
                      & $\hat\beta_1$ & 0.0016                & 0.0012                  & 0.0013                 & 0.0013                  & 0.0013                  \\
                      & $\hat\phi$   & 0.0094                & 0.0075                  & 0.0091                 & 0.0131                  & 0.0109                  \\ \hline
\end{tabular}
\label{tab: sens analysis one percent}
\end{table}

\begin{table}[H]
\centering
\caption{Bias and mean squared error (MSE) of the regression parameter estimates for the beta and BSM models when 5\% of the response values are replaced by atypical observations.}
\begin{tabular}{llrrrrr}
\hline
\multicolumn{1}{l}{}  &       & \multicolumn{5}{c}{5\%}                                                                                                      \\ \cline{3-7} 
\multicolumn{1}{l}{}  &       & \multicolumn{1}{c}{B} & \multicolumn{1}{c}{TPB} & \multicolumn{1}{c}{GB} & \multicolumn{1}{c}{LNB} & \multicolumn{1}{c}{IGB} \\ \hline
\multirow{3}{*}{bias} & $\hat\beta_0$ & -0.0427               & -0.0180                 & -0.0203                & -0.0218                 & -0.0228                 \\
                      & $\hat\beta_1$ & -0.0855               & -0.0369                 & -0.0413                & -0.0443                 & -0.0463                 \\
                      & $\hat\phi$   & 0.1251                & 0.0896                  & 0.1166                 & 0.1421                  & 0.1517                  \\
\multicolumn{1}{l}{}  &       & \multicolumn{1}{l}{}  & \multicolumn{1}{l}{}    & \multicolumn{1}{l}{}   & \multicolumn{1}{l}{}    & \multicolumn{1}{l}{}    \\
\multirow{3}{*}{MSE}  & $\hat\beta_0$ & 0.0031                & 0.0014                  & 0.0015                 & 0.0015                  & 0.0016                  \\
                      & $\hat\beta_1$ & 0.0091                & 0.0027                  & 0.0031                 & 0.0034                  & 0.0035                  \\
                      & $\hat\phi$   & 0.0158                & 0.0081                  & 0.0138                 & 0.0216                  & 0.0236                  \\ \hline
\end{tabular}
\label{tab: sens analysis five percent}
\end{table}
\section{Data Application}\label{sec:Data application}
In this section, the BSM regression model introduced in Section \ref{sec: Examples of UBSM} is applied to real-world data sets, namely the \texttt{MockJurors} and \texttt{sdac} datasets, which are available in the \texttt{betareg} and \texttt{simplexreg} packages in \texttt{R}, respectively. To illustrate the model's viability as an alternative to the beta regression model, we benchmark it to other models defined on the unit interval. These competing models include the beta rectangular (BR; \citealp{hahn2008mixture,bayes2012new}), Kumaraswamy (KW; \citealp{kumaraswamy1980generalized}), beta–Kumaraswamy (BKW), generalized Kumaraswamy (GKW; \citealp{carrasco2010new}), logit-normal (LogitN), and generalized beta type I (GB1) regression models. The GKW-based models were fitted using the \texttt{gkwreg} package \citep{lopes2026gkwreg}, whereas the LogitN and GB1 models were fitted using the \texttt{GAMLSS} package \citep{stasinopoulos2017flexible}.  Model performance is ranked via the Akaike information criterion (AIC; \citealp{akaike1974new}),
\begin{align*}
    \text{AIC}=2k-2\ell(\hat{\boldsymbol{\beta}},\hat{\phi},\hat{\boldsymbol{\theta}}),
\end{align*}
 and the Bayesian information criterion (BIC; \citealp{schwarz1978estimating}),
\begin{align*}
    \text{BIC} = \mathrm{log}(n)k-2\ell(\hat{\boldsymbol{\beta}},\hat{\phi},\hat{\boldsymbol{\theta}}) ,
\end{align*}
where $k$ is the number of parameters, $n$ is the number of observations, 
$\hat{\boldsymbol\beta},\hat{\phi}$, and $\hat{\boldsymbol{\theta}}$ are the ML estimates of $\boldsymbol{\beta}, \phi$ and $\boldsymbol{\theta}$, respectively, and where $\ell(\hat{\boldsymbol{\beta}},\hat{\phi},\hat{\boldsymbol{\theta}})$ is the maximized log-likelihood value for the BSM model.

\subsection{Mock jurors data}
The mock jurors dataset consists of 104 responses examining how jurors’ confidence in their verdict is affected by the use of a conventional two-option verdict (guilty vs. acquittal) compared with a three-option verdict system (including the Scottish “not proven” alternative), as well as by the presence or absence of conflicting testimonial evidence. 



The BSM regression model uses \texttt{confidence} as the response variable and \texttt{verdict} as the explanatory variable. Formally, the model is specified as
\begin{align*}
    \text{confidence}_i|\text{verdict}_i&\sim \mathcal{BSM}(\mu(\text{verdict}_i;\boldsymbol{\beta}),\phi,\,\boldsymbol{\theta})\\
    \mathrm{logit}(\mu(\text{verdict}_i;\boldsymbol{\beta}))&=\beta_0+\beta_1\text{verdict}_i
\end{align*}
for $i=1, \dots,104.$ Model comparison results based on AIC and BIC are summarized in Table~\ref{tab:Ranking_fitted_reg_MockJurors}. Among all competing models, the BSM regression models consistently outperform the standard beta and competitor models. In particular, the IGB regression model achieves the lowest AIC and BIC values, indicating superior overall fit.

\begin{table}[H]
\centering
\caption{Ranking of fitted regression models to mock jurors data according to AIC and BIC.} 
\begin{tabular}{lrrrrrr}
  \hline
 Model & \#par & log-likelihood & AIC & Rank & BIC & Rank  \\ 
  \hline
B &     3 & 28.5806 & -51.1611 &     8 & -43.2280 &     7 \\ 
  TPB &     5 & 38.9206 & -67.8411 &     4 & -54.6192 &     4 \\ 
  GB &     4 & 38.0714 & -68.1428 &     3 & -57.5652 &     3 \\ 
  IGB &     4 & 38.4982 & -68.9963 &     1 & -58.4188 &     1 \\ 
  LNB &     4 & 38.4423 & -68.8846 &     2 & -58.3071 &     2 \\ 
  BR &     4 & 35.7298 & -63.4596 &     5 & -52.8820 &     5 \\ 
  GKW &     6 & 29.8171 & -47.6342 &    10 & -31.7678 &    10 \\ 
  KW &     3 & 28.8426 & -51.6853 &     7 & -43.7521 &     6 \\ 
  BKW &     5 & 30.4264 & -50.8527 &     9 & -37.6308 &     9 \\ 
  LogitN &     3 & 22.0126 & -38.0252 &    11 & -30.0920 &    11 \\ 
  GB1 &     5 & 32.1201 & -54.2403 &     6 & -41.0183 &     8 \\ 
   \hline
\end{tabular}
\label{tab:Ranking_fitted_reg_MockJurors}
\end{table}


Table \ref{tab: estimates Stats MockJurors} presents the estimates of the beta and IGB regression models, along with their standard errors (SE) in parentheses. Here, it is clear that the standard errors are smaller for the IGB than for the beta regression model. Since a logit link function is used for the parameter $\mu$, the coefficients describe the change in log-odds of the outcome variable as a function of the jurors' verdict. 

For the beta regression model, the intercept is estimated at $\hat\beta_0=0.8521$ (SE = 0.1096), representing the baseline level of juror confidence on the link scale for the reference verdict category. The coefficient for verdict is 
$\hat\beta_1$=0.0859 (SE = 0.1019), indicating a small positive change in the transformed mean confidence when moving from the reference verdict to the alternative verdict system. This effect is relatively modest, and the standard error suggests some uncertainty in the estimate. The variability parameter is $\hat\phi=0.3854$ (SE = 0.3274), reflecting considerable variability in juror confidence under the beta model.

Under the IGB model, the intercept is slightly higher at 
$\hat\beta_0=0.9120$ (SE = 0.0972), indicating a higher baseline transformed mean confidence. The verdict coefficient increases to $\hat\beta_1=0.1424$ (SE = 0.0952), suggesting that the alternative verdict system is associated with a more pronounced increase in expected juror confidence compared with the beta regression. The corresponding variability parameter is $\hat\phi=0.6636$ (SE = 0.1998), indicating that the IGB regression model captures juror confidence with less unexplained variability. The additional shape parameter, 
$\hat\theta=8.0196$ (SE = 9.0492), allows the model to accommodate the observed skewness and heavy tails in the data, though the large standard error indicates some uncertainty in this estimate.

Overall, the IGB regression model yields slightly larger verdict effects and more precise coefficient estimates compared to the beta model, reflecting a better fit to the skewed and heavy-tailed distribution of juror confidence.
\begin{table}[H]
\caption{Estimated coefficients of fitted regression models to Mock Jurors data.} \label{tab: estimates Stats MockJurors}
\centering
\begin{tabular}{lrrr}
\hline
Parameter & \multicolumn{1}{c}{B} & \multicolumn{1}{c}{IGB} \\ \hline
$\hat\beta_0$ & 0.8521 (0.1096)   & 0.9120 (0.0972)    \\
$\hat\beta_1$   & 0.0859 (0.1019)   & 0.1424 (0.0952)    \\
$\hat\phi$       & 0.3854 (0.3274)   & 0.6636 (0.1998)    \\
$\hat\theta$     &                      & 8.0196  (9.0492)    \\ \hline
\end{tabular}
\end{table}

\subsection{Autologous peripheral blood stem cell transplants data}

Autologous peripheral blood stem cell (PBSC) transplantation is widely used to restore hematologic function after myeloablative chemotherapy. Successful engraftment depends on reinfusing sufficient numbers of viable CD34+ stem cells, but the processes of freezing, cryopreservation, and thawing can damage cells and reduce viability.

The \texttt{sdac} dataset used in \cite{zhang2016simplexreg} arises from a study investigating factors that influence post-cryopreservation recovery rates of viable CD34+ cells. It includes $n=242$ patients who underwent autologous PBSC transplantation between 2003 and 2008 at the Edmonton Hematopoietic Stem Cell Laboratory. The response variable, \texttt{rcd}, is the recovery rate of CD34+ cells, defined as the percentage of viable CD34+ cells measured after thawing relative to the number measured before freezing.

Explanatory variables include patient age and chemotherapy protocol. Age is represented by \texttt{ageadj}, a continuous variable measured in years and defined as patient age minus 40, with age 40 serving as the baseline. Chemotherapy protocol is represented by \texttt{chemo}, a binary indicator coded as 0 for a 1-day protocol and 1 for a 3-day protocol, with other regimens grouped according to their duration.

The model is specified as
\begin{align*}
    \text{rcd}_i|\text{ageadj}_i,\text{chemo}_i&\sim \mathcal{BSM}(\mu(\text{ageadj}_i,\text{chemo}_i;\boldsymbol{\beta}),\phi,\boldsymbol{\theta})\\
    \mathrm{logit}(\mu(\text{ageadj}_i,\text{chemo}_i;\boldsymbol{\beta}))&=\beta_0+\beta_1\text{ageadj}_i+\beta_2\text{chemo}_i
\end{align*}
for $i=1, \dots,242$.

\begin{table}[H]
\centering
\caption{Ranking of fitted regression models to Autologous Peripheral Blood Stem Cell Transplants Data according to AIC and BIC.} 

\begin{tabular}{lrrrrrr}
  \hline
Model & \#par & log-likelihood & AIC & Rank & BIC & Rank \\ 
  \hline
B &   4 & 195.65 & -383.30 &   7 & -369.40 &   4 \\ 
  TPB &   6 & 200.64 & -389.27 &   4 & -368.41 &   6 \\ 
  GB &   5 & 199.68 & -389.37 &   3 & -371.99 &   3 \\ 
  IGB &   5 & 200.10 & -390.20 &   1 & -372.81 &   1 \\ 
  LNB &   5 & 200.00 & -390.00 &   2 & -372.62 &   2 \\ 
  BR &   5 & 195.55 & -381.11 &   9 & -363.73 &   8 \\ 
  GKW &   7 & 200.21 & -386.42 &   5 & -362.09 &   9 \\ 
  KW &   4 & 195.40 & -382.79 &   8 & -368.89 &   5 \\ 
  BKW &   6 & 199.04 & -386.08 &   6 & -365.22 &   7 \\ 
  LogitN &   4 & 182.61 & -357.22 &  11 & -343.32 &  11 \\ 
  GB1 &   6 & 195.53 & -379.07 &  10 & -358.21 &  10 \\ 
   \hline
\end{tabular}
\label{tab:Ranking_fitted_reg_data}
\end{table}

Among all competing models, the BSM regression models outperform the standard beta and other competitor models. In particular, the IGB regression model achieves the lowest AIC and BIC values again, indicating the best overall fit to the post-cryopreservation CD34+ cell recovery data (Table \ref{tab:Ranking_fitted_reg_data}). Notably, the TPB model does not rank among the top models according to the BIC, highlighting that the IGB, LNB, and GB regression models capture the data more efficiently when accounting for model complexity. This emphasizes the additional flexibility of the BSM regression model.

Table \ref{tab: estimates Stats PBSC} presents the estimated coefficients of the beta and IGB regression models, along with their standard errors in parentheses. Overall, the IGB model yields slightly more precise regression coefficient estimates than the standard beta model. 

\begin{table}[H]
\caption{Estimated coefficients of fitted regression models to Autologous Peripheral Blood Stem Cell Transplants data.} \label{tab: estimates Stats PBSC}
\centering
\begin{tabular}{lrr}
\hline
Parameter     & \multicolumn{1}{c}{B} & \multicolumn{1}{c}{IGB} \\ \hline
$\hat\beta_0$ & 1.0422 (0.1115)       & 1.0171 (0.1096)         \\
$\hat\beta_1$ & 0.0143 (0.0053)       & 0.0159 (0.0052)         \\
$\hat\beta_2$ & 0.2143 (0.1017)       & 0.1876 (0.0985)         \\
$\hat\phi$    & 0.0883 (0.0079)       & 0.1190 (0.0273)         \\
$\hat\theta$  &                       & 1.1071 (1.0816)         \\ \hline
\end{tabular}
\end{table}

For the beta regression model, the intercept is estimated at $\hat\beta_0=1.0422$ (SE = 0.1115), representing the baseline log-odds of CD34+ cell recovery for a 40-year-old patient on the 1-day chemotherapy protocol. The age effect is small but positive, with $\hat\beta_1=0.0143$ (SE = 0.0053), indicating a slight increase in expected recovery per year above age 40. The effect of chemotherapy protocol is more substantial, with $\hat\beta_2=0.2143$ (SE = 0.1017), suggesting that patients receiving the 3-day protocol tend to have higher post-thaw recovery rates. The variability parameter is $\hat\phi=0.0883$ (SE = 0.0079), indicating some variability in recovery rates across patients.

Under the IGB model, the intercept is slightly lower at $\hat\beta_0=1.0171$ (SE = 0.1096), and the age coefficient increases marginally to $\hat\beta_1=0.0159$ (SE = 0.0052), reflecting a slightly stronger age-related increase in recovery. The chemotherapy protocol effect is somewhat smaller, $\hat\beta_2=0.1876$ (SE = 0.0985), but remains positive. The precision parameter is higher than in the beta model, $\hat\phi=0.1190$ (SE = 0.0273), indicating that the IGB model captures recovery with less unexplained variability. The additional shape parameter, $\hat\theta=1.1071$ (SE = 1.0816), allows the model to account for skewness in the distribution of recovery rates, although the relatively large standard error reflects some uncertainty in this estimate.

Overall, the IGB regression model provides slightly more precise and robust estimates of the effects of age and chemotherapy protocol on post-thaw CD34+ cell recovery, confirming its better fit to the skewed and variable distribution observed in this clinical dataset. 

\section{Conclusion} \label{Sec:conclusion}
With the advancement of information technology, the prevalence of skewed, leptokurtic, and heavy-tailed data has increased substantially, particularly across domains such as finance, engineering, and medicine. For data supported on the unit interval, the beta distribution is traditionally regarded as the standard choice. However, this paper shows that the proposed BSM model provides a competitive alternative, demonstrating favorable performance relative to several existing bounded-domain models. Nevertheless, the contribution of the present study extends beyond the four illustrative models considered, and the applicability of the proposed model is not confined to the datasets considered. In particular, additional mixture distributions may be readily incorporated, and the proposed framework can be adapted to a wide range of contexts for the unit interval.

A natural and practically relevant extension of the BSM regression model is to allow all model parameters to depend on covariates, thereby substantially enhancing its flexibility and broadening its applicability in regression settings.

\section*{Acknowledgments}
Ferreira has been partially supported by: (i) the National Research Foundation (NRF) of South Africa (SA), grant RA201125576565, nr. 145681 and CSRP250220299317; and (ii) the DSI-NRF Centre of Excellence in Mathematical and Statistical Sciences (CoE-MaSS). 
Bekker acknowledges the support of the National Research Foundation (NRF) of South Africa (SA), grant RA231117164450. 
The opinions expressed and conclusions arrived at are those of the authors and are not necessarily to be attributed to the NRF.

\section*{Data availability statement}
All datasets considered in this paper are freely available in the \texttt{betareg} and \texttt{simplexreg} packages in \texttt{R}. All code used for the simulations and real data applications was developed and executed in RStudio \citep{RStudio} (Version 2025.05.1+513).

\section*{Disclosure statement}
The authors declared no potential conflicts of interest with respect to the research, authorship, and/or publication of this article.







\bibliographystyle{chicago}
\bibliography{database}

\end{document}